\documentclass[sigconf,10pt,sort, nonacm]{acmart}

\renewcommand\footnotetextcopyrightpermission[1]{}

\AtBeginDocument{%
  \providecommand\BibTeX{{%
    \normalfont B\kern-0.5em{\scshape i\kern-0.25em b}\kern-0.8em\TeX}}}

\usepackage{amsmath}
\usepackage{soul}
\usepackage{enumitem}
\usepackage{multirow}
\usepackage{booktabs}
\usepackage{array}
\usepackage{pifont}
\usepackage{tikz}
\usepackage{pgfplots}
\usepackage{algpseudocode}
\usepackage{algorithm}
\usepackage{algorithmicx}
\usepackage{algpseudocode}
\usepackage{csquotes}

\usepackage{tcolorbox}
\usepackage{amsmath, bm}
\usepackage{harpoon}

\usepackage[colorinlistoftodos]{todonotes}

\usepackage[subrefformat=parens, labelformat=parens]{subfig}
\usepackage[font=small]{caption}
\usepackage{listings}
\definecolor{darkgreen}{rgb}{0, 0.5, 0}
\usepgfplotslibrary{external}

\usepackage{atbegshi} 

\newcommand{\acceptedbox}{%
  \begin{tcolorbox}[
    colback=gray!10,
    colframe=black,
    boxrule=0.5pt,
    arc=0pt,
    width=\textwidth,
    left=6pt,right=6pt,top=4pt,bottom=4pt
  ]
  \small
  This paper has been accepted to the \textbf{30th Annual International Conference on Mobile Computing and Networking (ACM MobiCom 2026).}\\[-1pt]
  Copyright may be transferred without notice, after which this version may no longer be accessible.
  \end{tcolorbox}%
}

\begin{document}

\AtBeginShipoutFirst{%
  \AtBeginShipoutUpperLeft{%
    \vbox to 0pt{%
      \vspace*{0.6in} 
      \hbox to \paperwidth{\hspace*{0.75in}\acceptedbox\hss}%
      \vss
    }%
  }%
}

\setlength{\abovedisplayskip}{3pt}
\setlength{\belowdisplayskip}{3pt}

\thispagestyle{plain}

\definecolor{lightgray}{gray}{0.9}
\definecolor{lightblue}{rgb}{0.9,0.9,1}
\definecolor{LightMagenta}{rgb}{1,0.5,1}
\definecolor{red}{rgb}{1,0,0}
\definecolor{brightgreen}{rgb}{0.4, 1.0, 0.0}

\newcommand\couldremove[1]{{\color{lightgray} #1}}
\newcommand{\remove}[1]{}
\newcommand{\move}[2]{ {\textcolor{Purple}{ \bf --- MOVE #1: --- }} {\textcolor{Orchid}{#2}} }

\newcommand{\hlc}[2][yellow]{ {\sethlcolor{#1} \hl{#2}} }
\newcommand\note[1]{\hlc[SkyBlue]{-- #1 --}} 

\newcommand\mynote[1]{\hlc[yellow]{#1}}
\newcommand\tingjun[1]{\hlc[yellow]{TC: #1}}
\newcommand\zhihui[1]{\hlc[LightMagenta]{ZG: #1}}
\newcommand\tom[1]{\hlc[brightgreen]{TOM: #1}}
\newcommand\andy[1]{\hlc[pink]{Andy: #1}}
\newcommand\change[1]{{\color{blue} {#1}}}

\newcommand{\TODO}[1]{\textcolor{red}{#1}}
\newcommand{\revise}[1]{\textcolor{blue}{#1}}


\newcommand{\myparatight}[1]{\vspace{0.5ex}\noindent\textbf{#1~~}}

\newcommand{\cmark}{\ding{51}}%
\newcommand{\xmark}{\ding{55}}%
\newcommand{\greencheck}{\color[HTML]{3C8031}{\cmark}}
\newcommand{\redcross}{\color[HTML]{ED1B23}{\xmark}}
\newcommand{\greenno}{\color[HTML]{3C8031}{\textbf{No}}}
\newcommand{\redyes}{\color[HTML]{ED1B23}{\textbf{Yes}}}
\newcommand{\greenlow}{\color[HTML]{3C8031}{\textbf{Low}}}
\newcommand{\redhigh}{\color[HTML]{ED1B23}{\textbf{High}}}

\newcommand*\circledwhite[1]{\tikz[baseline=(char.base)]{
            \node[shape=circle,draw,inner sep=0.6pt] (char) {\scriptsize{#1}};}}

\newcommand*\circled[1]{\tikz[baseline=(char.base)]{
            \node[shape=circle,draw,fill=black,text=white,inner sep=0.6pt] (char) {\scriptsize{#1}};}}

\newcommand{\name}{{\sf Nexus}}
\newcommand{\namebf}{{\sf\textbf{Nexus}}}

\newcommand{\namepdo}{{\sf\textbf{PDO}}}
\newcommand{\namesddo}{{\sf\textbf{DDO}}}
\newcommand{\namesmlo}{{\sf\textbf{MLO-X}}}
\newcommand{\namesfwo}{{\sf\textbf{FWO}}}

\newcommand{\namesc}{{\sf\textbf{Savannah-sc}}}
\newcommand{\namemc}{{\sf Savannah-mc}}

\newcommand{\savannah}{Savannah}
\newcommand{\savannahsc}{Savannah-sc}
\newcommand{\savannahmc}{Savannah-mc}
\newcommand{\agora}{Agora}
\newcommand{\agorabf}{\textbf{Agora}}

\newcommand{\armavec}{\sf Savannah-mc (arma-vec)}
\newcommand{\armacube}{\sf Savannah-mc (arma-cube)}

\newcommand{\specialcell}[2][c]{%
  \begin{tabular}[#1]{@{}c@{}}#2\end{tabular}}

\newcommand{\iu}{{j}}

\newcommand{\littlesum}{\mathop{\textstyle\sum}}
\newcommand{\littleint}{\mathop{\textstyle\int}}

\newcommand{\siso}{SISO}
\newcommand{\mimoTwoByTwo}{2$\times$2}
\newcommand{\mimoFourByFour}{4$\times$4}

\newcommand{\myCodeShort}[1]{\texttt{\small{#1}}}

\newcommand{\bbdev}{\textsf{bbdev}}

\newcommand{\scs}{{\Delta f}}
\newcommand{\scNum}{N_{\textrm{sc}}}
\newcommand{\sampRate}{F_{\textrm{s}}}
\newcommand{\fftSize}{N_{\textrm{FFT}}}
\newcommand{\chMat}{{\textbf{H}}}
\newcommand{\chVec}{{\textbf{h}}}
\newcommand{\precodeMat}{{\textbf{P}}}

\newcommand{\usec}{$\upmu$s} 
\newcommand{\msec}{ms}       

\newcommand{\fft}{\textsf{fft}}
\newcommand{\ifft}{\textsf{ifft}}
\newcommand{\csi}{\textsf{csi}}
\newcommand{\precode}{\textsf{precode}}
\newcommand{\encode}{\textsf{enc}}
\newcommand{\decode}{\textsf{dec}}
\newcommand{\modul}{\textsf{modul}}
\newcommand{\demod}{\textsf{demod}}
\newcommand{\equal}{\textsf{equal}}

\newcommand{\tbSize}{T}
\newcommand{\tbCrcSize}{T_{\textrm{crc}}}
\newcommand{\cbSize}{K_{\textrm{cb}}}
\newcommand{\cbNum}{N_{\textrm{cb}}}
\newcommand{\liftingSize}{Z_{c}}
\newcommand{\liftingSizeSet}{\mathbf{\Theta}}
\newcommand{\fillerBitNum}{N_{\textrm{filler}}}

\newcommand{\codeRate}{R}

\newcommand{\throughput}{Tp}
\newcommand{\codingTime}{t}
\newcommand{\informationBits}{K'}

\newcommand{\latency}{\mathcal{L}}
\newcommand{\power}{P}

\newcommand{\cellConfigMIMO}{M}
\newcommand{\cellConfigBW}{B}
\newcommand{\cellConfigTL}{\rho_{t}}
\newcommand{\cellConfigTBW}{\rho_{f}}
\newcommand{\cellConfigMCS}{\chi}
\newcommand{\cellConfigVar}{\bm{\upsigma}}
\newcommand{\cellConfig}[5]{\bm{\upsigma}({#1}, {#2}, {#3}, {#4}, {#5})}
\newcommand{\cellProfile}{\mathcal{P}}

\newcommand{\sysConfigVar}{\bm{\uptheta}}
\newcommand{\sysConfig}[4]{\bm{\uptheta}({#2}, {#3} | {#1}, {#4})}
\newcommand{\sysConfigOpt}{\bm{\uptheta}^{\star}}

\newcommand{\setSysConfigVar}{\bm{\upTheta}}
\newcommand{\setSysConfig}[2]{\bm{\upTheta}({#1}, {#2})}
\newcommand{\setSysConfigReduced}[2]{\bm{\upTheta}'({#1}, {#2})}

\newcommand{\sysConfigCellIdx}[5]{\bm{\uptheta}_{#1}({#3}, {#4} | {#2}, {#5})}

\newcommand{\setSize}[1]{|{#1}|}

\newcommand{\numCoreMax}{C_{\textrm{max}}}
\newcommand{\setCoreTotal}{\mathcal{C}}
\newcommand{\numCoreTotal}{C}
\newcommand{\setCoreTotalCellIdx}[1]{\mathcal{C}_{#1}}
\newcommand{\numCoreTotalCellIdx}[1]{C_{#1}}

\newcommand{\setCoreDsp}{\mathcal{C}^{\textrm{dsp}}}
\newcommand{\numCoreDsp}{C^{\textrm{dsp}}}
\newcommand{\setCoreDspCellIdx}[1]{\mathcal{C}_{#1}^{\textrm{dsp}}}
\newcommand{\numCoreDspCellIdx}[1]{C_{#1}^{\textrm{dsp}}}

\newcommand{\setCoreAcc}{\mathcal{C}^{\textrm{acc}}}
\newcommand{\numCoreAcc}{C^{\textrm{acc}}}
\newcommand{\setCoreAccCellIdx}[1]{\mathcal{C}_{#1}^{\textrm{acc}}}
\newcommand{\numCoreAccCellIdx}[1]{C_{#1}^{\textrm{acc}}}

\newcommand{\numVfAccMax}{V_{\textrm{max}}}
\newcommand{\setVfAcc}{\mathcal{V}}
\newcommand{\numVfAcc}{V}
\newcommand{\setVfAccCellIdx}[1]{\mathcal{V}_{#1}}
\newcommand{\numVfAccCellIdx}[1]{V_{#1}}

\newcommand{\costRatioDspAcc}{\alpha}

\newcommand{\myAbs}[1]{\left|{#1}\right|}
\newcommand{\myAng}[1]{\angle{#1}}
\newcommand{\myConjugate}[1]{{#1}^{*}}
\newcommand{\myTranspose}[1]{{#1}^{\top}}
\newcommand{\myHermitian}[1]{{#1}^{H}}
\newcommand{\myIsFunc}[1]{\mathbf{1}\{#1\}}

\newcommand{\AoD}{\phi}
\newcommand{\AoDVec}{\bm{\upphi}}
\newcommand{\AoDbf}{\boldsymbol\phi}
\newcommand{\AoDDirectional}{\Phi}
\newcommand{\az}{\phi}
\newcommand{\azVec}{\bm{\upphi}}
\newcommand{\azVecUE}{\bm{\upphi}_{\textrm{UE}}}
\newcommand{\azbf}{\boldsymbol\phi}
\newcommand{\el}{\psi}
\newcommand{\elbf}{\boldsymbol\psi}

\newcommand{\ElemComp}{w}
\newcommand{\ElemCompbf}{\mathbf{w}}
\newcommand{\ElemCompNew}{w^\prime}
\newcommand{\ElemCompNewbf}{\mathbf{w}^\prime}
\newcommand{\ElemAmp}{A}
\newcommand{\ElemAmpbf}{\mathbf{A}}
\newcommand{\ElemPhase}{\theta}
\newcommand{\ElemPhasebf}{\boldsymbol\theta}
\newcommand{\steer}{s}
\newcommand{\steerVec}{\mathbf{s}}
\newcommand{\steermat}{\mathbf{S}}
\newcommand{\beamPattern}{BP}

\newcommand{\bw}{B}
\newcommand{\carrierFreq}{f_{c}}
\newcommand{\carrierWave}{\lambda}

\newcommand{\csiMat}{\mathbf{H}}

\newcommand{\ASA}[2]{\textrm{ASA}({#1},{#2})}
\newcommand{\antNum}{N}
\newcommand{\antIdx}{n}
\newcommand{\antDist}{d}
\newcommand{\subarrayNum}{M}
\newcommand{\subarraySet}{\mathcal{M}}
\newcommand{\subarrayIdx}{m}
\newcommand{\subarrayAntNum}{N_{s}}
\newcommand{\subarrayAntIdx}{n}
\newcommand{\subarrayAntDist}{d}

\newcommand{\setSubarray}{\mathcal{A}}
\newcommand{\subarrayAlloc}{a}
\newcommand{\subarrayAllocVec}{\mathbf{a}}
\newcommand{\subarrayAllocMat}{\mathbf{A}}
\newcommand{\subarrayAllocSet}{\mathbb{A}}

\newcommand{\bfWeight}{w}
\newcommand{\bfWeightVec}{\mathbf{w}}
\newcommand{\bfAmp}{A}
\newcommand{\bfAmpVec}{\mathbf{A}}
\newcommand{\bfPhase}{\theta}
\newcommand{\bfPhaseVec}{\boldsymbol{\theta}}
\newcommand{\bfGain}{g}
\newcommand{\bfGainSig}[1]{g^{\textrm{sig}}_{#1}}
\newcommand{\bfGainInt}[2]{g^{\textrm{int}}_{{#1}\rightarrow{#2}}}


\newcommand{\userNum}{U}
\newcommand{\userIdx}{u}
\newcommand{\userSet}{\mathcal{U}}

\newcommand{\userNumSub}{K}

\newcommand{\cellFeasibleRF}[1]{\mathcal{F}_{\mathrm{RAF}}\!\left(#1\right)}
\newcommand{\cellProfileRF}[1]{\mathcal{P}_{\mathrm{RAF}}\!\left(#1\right)}

\newcommand{\userSelected}{k}
\newcommand{\userSelectedNum}{K}
\newcommand{\userSelectedSet}{\mathcal{K}}

\newcommand{\userAngle}{\phi}
\newcommand{\userWeight}{\alpha}

\newcommand{\baseSNR}{\gamma}
\newcommand{\SNR}{\mathsf{SNR}}
\newcommand{\SNRMax}{\mathsf{SNR}^{\textrm{max}}}
\newcommand{\SINR}{\mathsf{SINR}}
\newcommand{\SINRMax}{\mathsf{SINR}^{\textrm{max}}}
\newcommand{\Capacity}{T}
\newcommand{\Rate}{R}
\newcommand{\RateMax}{\Rate^{\textrm{max}}}
\newcommand{\RateAvg}{\widebar{\Rate}}
\newcommand{\CapacityMax}{\Tilde{T}}
\newcommand{\suppress}{\alpha}

\newcommand{\RateHist}{\widebar{\Rate}}

\newcommand{\past}{p}
\newcommand{\decay}{\beta}

\newcommand{\RateMean}{\Bar{R}}
\newcommand{\JFI}{\mathsf{JFI}}

\newcolumntype{+}{>{\global\let\currentrowstyle\relax}}
\newcolumntype{^}{>{\currentrowstyle}}
\newcommand{\rowstyle}[1]{%
  \gdef\currentrowstyle{#1}%
  #1\ignorespaces
}

\newenvironment{spmatrix}[1]
 {\def\mysubscript{#1}\mathop\bgroup\begin{bmatrix}}
 {\end{bmatrix}\egroup_{\textstyle\mathstrut\mysubscript}}


\title[]{{\namebf}: Efficient and Scalable Multi-Cell mmWave Baseband Processing with Heterogeneous Compute}


\author{Zhenzhou Qi, Chung-Hsuan Tung, Zhihui Gao, Tingjun Chen}
\affiliation{%
  \institution{\vspace{0.5ex} Department of Electrical and Computer Engineering, Duke University\vspace{0.5ex}}
  \country{}
}

\renewcommand{\shortauthors}{~}

\begin{abstract}
The rapid adoption of 5G New Radio (NR), particularly in the millimeter-wave (mmWave) spectrum, imposes stringent demands on the flexibility, scalability, and efficiency of baseband processing.
While virtualized Radio Access Networks (vRANs) enable dynamic spectrum sharing across cells, compute resource allocation for baseband processing, especially in multi-cell deployments with heterogeneous workloads, remains underexplored.
In this paper, we present {\name}, the first system to realize real-time, virtualized multi-cell mmWave baseband processing on a single server with heterogeneous compute resources.
{\name} integrates software-based digital signal processing pipelines with hardware-accelerated LDPC decoding, and introduces a novel framework for sharing Intel's ACC100 eASIC across multiple CPU cores via virtual functions (VFs).
For single-cell operation, {\name} employs a random forest (RAF)-based model that predicts the most energy-efficient resource allocation for the given cell configuration with microsecond-level inference latency and high accuracy.
For multi-cell scenarios, {\name} introduces a power-aware scheduler that incorporates a lightweight contention model to adjust resource allocation strategies under concurrent execution.
Through extensive evaluation across various Frequency Range 2 (FR2) cell configurations, we show that {\name} supports up to 16 concurrent cells under full load, achieving {5.37}\thinspace{Gbps} aggregate throughput, while reducing the multi-cell scheduling search space by orders of magnitude.
These results demonstrate that virtualized, resource-aware baseband processing is both practical and efficient for next-generation vRAN systems.
\end{abstract}

\maketitle

\section{Introduction}
\label{sec:intro}







5G New Radio (NR) has rapidly evolved to become the foundation for modern wireless communication systems, aiming to deliver ultra-high throughput, ultra-low latency, and massive device connectivity~\cite{3gpp38.801}. A key feature of 5G NR is its flexible subcarrier spacings and slot durations to adapt to a variety of deployment scenarios. Combined with large bandwidth support and operation across diverse frequency ranges, including the millimeter-wave (mmWave) spectrum in Frequency Range 2 (FR2)~\cite{3gpp38.101-1}, 5G NR is pushing the limits of both radio access and infrastructure design~\cite{ rangan2014millimeter, rappaport2013millimeter, hunt2023madradar}.

To meet growing performance demands, the wireless industry is transitioning from dedicated hardware toward centralized Radio Access Network (RAN)~\cite{fan2018enabling, wu2015cloud, chih2014recent}, and further into disaggregated, software-defined architectures such as virtualized RAN (vRAN)~\cite{foukas2021concordia, qi2024savannah, garcia2018fluidran} and Open RAN (O-RAN)~\cite{singh2020evolution, wani2024open}. 
These architectures decouple baseband functions from specialized hardware, allowing deployment on general-purpose compute platforms with heterogeneous compute resources, including CPU, GPU, FPGA, and ASICs.
This shift reduces capital expenditure and enables key capabilities such as dynamic network slicing, centralized scheduling, coordinated beamforming, and elastic scaling of compute resources across diverse baseband workloads.

While significant research and development have focused on the virtualization of radio resources, e.g., dynamic Physical Resource Block (PCB) sharing~\cite{an2023deep}, little attention has been given to sharing computational resources in vRAN.
These computational resources, such as CPU cores, hardware accelerators (e.g., ASICs), and memory bandwidth, are just as critical as radio spectrum in determining the system's capacity, performance, and responsiveness.
As 5G moves toward dense, high-throughput deployments, the ability to efficiently share and schedule computational resources across cells becomes essential.

\begin{figure}[!t]
    \centering    
    \includegraphics[width=0.99\columnwidth]{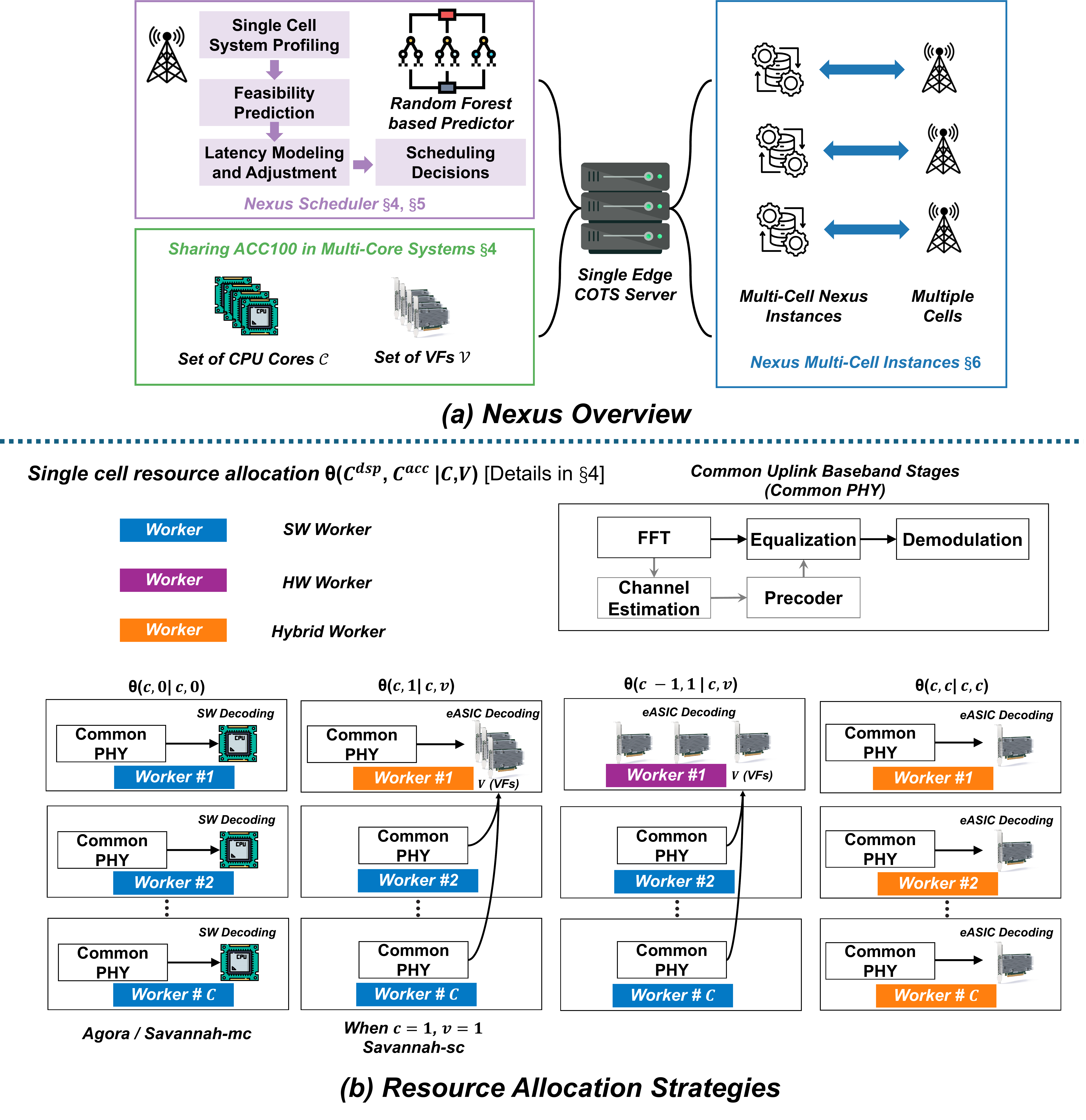}
    \vspace{-3mm}
    \caption{Overview of {\namebf}: 
    (a) {\namebf} virtualizes multi-cell mmWave PHY processing on a single server with heterogeneous compute by sharing a set of CPU cores and a pool of ACC100 virtual functions (VFs) in a power-aware manner.
    (b) Resource allocation strategies for a single cell are characterized by combinations of CPU cores and VFs. 
    }
    \label{fig:System-Design}
    \vspace{-5mm}
\end{figure}

In this paper, we present {\name}, the first system that enables efficient, virtualized multi-cell mmWave baseband processing on a single server with heterogeneous compute resources, including CPUs and ASICs. 
{\name} addresses the key challenge of scaling computational resources across multiple FR2 cells while guaranteeing the strict real-time processing deadlines imposed by 5G NR.  
Designing {\name} introduces several technical challenges.  
First, uplink Physical layer (PHY) processing in FR2 with numerology 3 and channel bandwidths up to {400}\thinspace{MHz} imposes extreme computational demand, especially under full traffic load.  
Second, integrating hardware accelerators such as Intel’s ACC100 eASIC for LDPC decoding into a shared multi-core vRAN environment requires careful coordination between CPU-based digital signal processing (DSP) stages and accelerator offloading, where queueing and PCIe transfer overheads can create unpredictable bottlenecks.  
Third, efficiently supporting heterogeneous cells requires a resource allocation framework that can (\emph{i}) adapt allocations to diverse PHY layer parameters, and (\emph{ii}) account for latency contention effects when multiple cells run concurrently on the same server.  

To overcome these challenges, {\name} introduces a complete system architecture (Fig.~\ref{fig:System-Design}).
It features the first detailed design for sharing ACC100 accelerators across multiple worker cores via virtual functions; a random forest (RAF)-based model that maps PHY-layer features, or \emph{cell configurations}, to the most energy-efficient resources required for each cell; and a power-aware scheduler for both single-cell and multi-cell scenarios, the latter integrating a lightweight contention model to ensure per-cell latency guarantees under concurrent execution.  
We extensively evaluate {\name} across various deployment scenarios.  
At the model level, we characterize the performance of our RAF-based predictor in detail.  
We compare it against six baselines, including three deep learning (DL) models and three analytical approaches, showing that RAF achieves near-DL accuracy with two orders of magnitude faster inference, smaller latency variance, and modest model size.  
We further demonstrate that the RAF model generalizes well to unseen single-cell configurations, sustaining $>98\%$ prediction accuracy.  
At the system level, we validate {\name} under diverse deployment scenarios.  
We show that {\name} supports up to sixteen concurrent cells with a total data rate of {5.37}\thinspace{Gbps} using sixteen CPU cores and sixteen ACC100 VFs, with the ability to scale further.

In summary, this paper makes the following contributions: 
\begin{itemize}[leftmargin=*, topsep=2pt, itemsep=1pt]
\item
We present {\name}, the first system supporting real-time, virtualized multi-cell mmWave baseband processing on a single server with heterogeneous compute resources (CPUs and ASICs).
This includes the first ACC100 sharing framework across multiple cores, enabling flexible offloading modes and scalable LDPC decoding;
\item
We design an RAF-based model that maps cell configurations to the most energy-efficient resource allocation required to meet the PHY processing deadline. 
For multi-cell deployments, we develop a contention-aware extension of the RAF scheduler that captures system-level compute resource contention across cells;
\item
We extensively evaluate {\name} on a heterogeneous compute testbed, demonstrating robust support for up to 16 concurrent FR2 cells at 100\% traffic load, achieving over {5.37}\thinspace{Gbps} aggregate throughput while meeting the 3-slot PHY processing deadline.
\end{itemize}

\noindent
The codebase and measurements of {\name} are open-sourced~\cite{NEXUS-github}.

\section{Related Work}

\myparatight{mmWave testbeds}
mmWave communication has emerged as a key enabler for high-throughput wireless systems, with its expansive spectrum in the {24–52}\thinspace{GHz} range accommodating higher data rates than {sub-6}\thinspace{GHz} technologies. 
The unique propagation characteristics in mmWave bands, such as higher path loss and susceptibility to blockage, have spurred research into advanced beamforming and beam management techniques that counteract short coverage ranges and ensure robust connectivity~\cite{woodford2021spacebeam, gao2024mambas}. To evaluate these strategies in realistic settings, a number of testbed~\cite{raychaudhuri2020challenge, qi2023programmable} have been proposed, including, programmable phased-array antennas~\cite{sadhu201728}, and prototype 5G NR FR2 equipment~\cite{zhao2020m}. 

\myparatight{PHY Baseband Processing and Virtualization}
Traditional baseband processing relies on specialized hardware accelerators or Digital Signal Processors (DSPs) capable of handling computationally intensive tasks. The advancement of virtualization techniques and cloud-native paradigms such as Sora~\cite{tan2011sora}, BigStation~\cite{yang2013bigstation}, Agora~\cite{ding2020agora}, Hydra~\cite{gong2023scalable} and Savannah~\cite{qi2024savannah} have driven a shift toward software-implemented PHY stacks on commodity hardware, often hosted on general-purpose CPUs, GPUs, or specialized accelerators. This virtualization of baseband functions allows network operators to allocate and scale resources based on changing load conditions, while also enabling easier deployment of updates and new features. 
To the best of our knowledge, our work is the first to specifically target virtualized multi-cell baseband processing in a single server, pushing beyond the typical single-cell or multi-server setups explored in prior literature.

\myparatight{Heterogeneous Computing and Resource Sharing}
Recent interest in heterogeneous computing has spurred work on efficiently utilizing diverse processing elements~\cite{wang2024understanding, schiavo2024cloudric}, such as CPUs, GPUs, and FPGAs, to tackle the demanding computational requirements of next-generation wireless networks. 
While many studies have extensively explored radio resource sharing, for example, dynamic allocation of physical resource blocks (PRBs) among different users~\cite{an2023deep}, relatively few have focused on computational resource sharing, especially in environments where multiple heterogeneous processors coexist within the same physical infrastructure.
This gap highlights a pressing need for strategies that can effectively partition, schedule, and manage such resources without compromising performance or scalability in high-frequency, high-bandwidth scenarios.

\section{Preliminaries}
\label{sec:prelim}

\begin{figure}[!t]
    \centering
    \vspace{-3mm}
    \subfloat[\hspace{1mm}$\numVfAcc \leq \numCoreTotal$]{
    \includegraphics[height=1.7in]{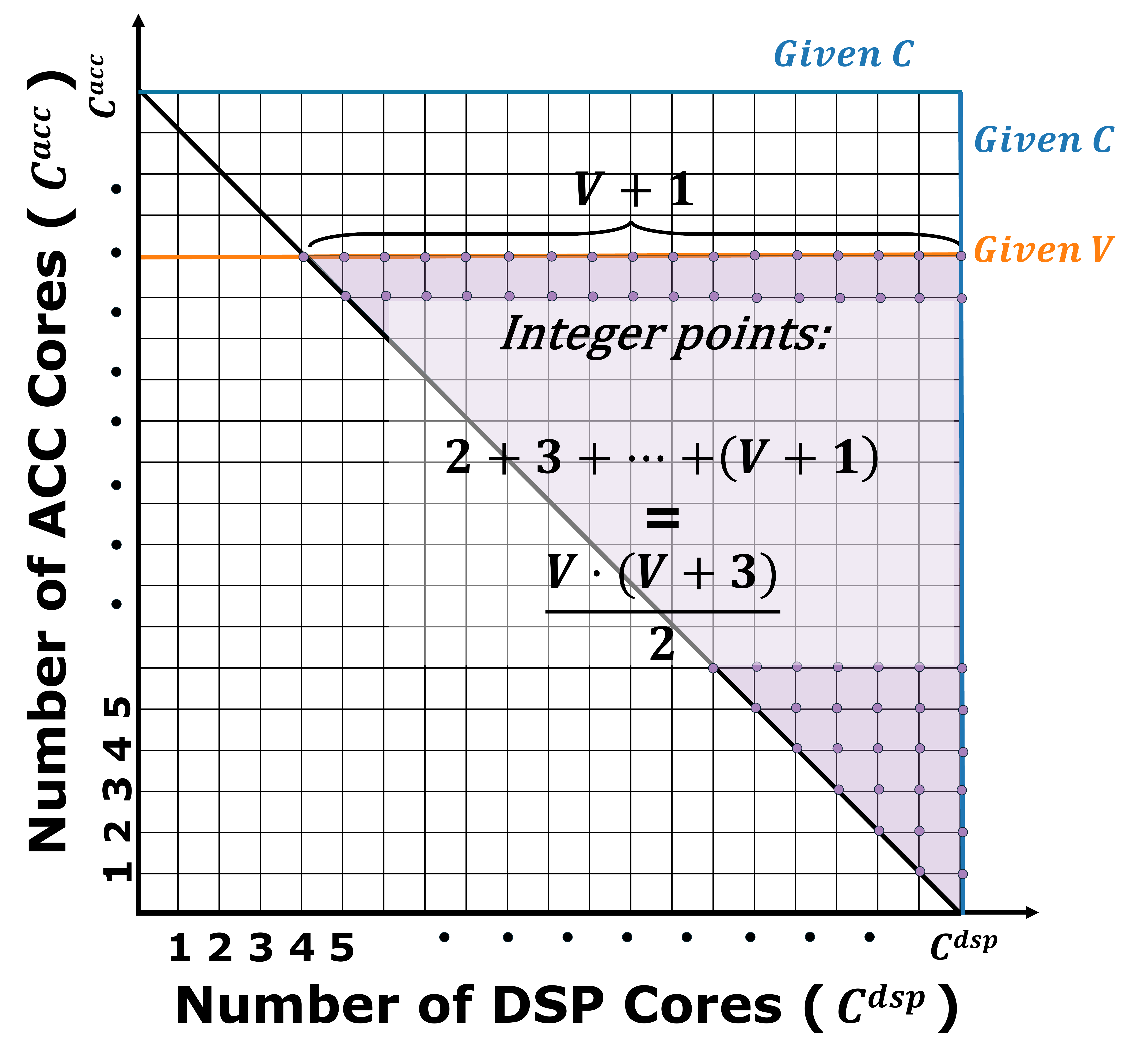}}
    \subfloat[\hspace{1mm}$\numVfAcc > \numCoreTotal$]{
    \includegraphics[height=1.7in]{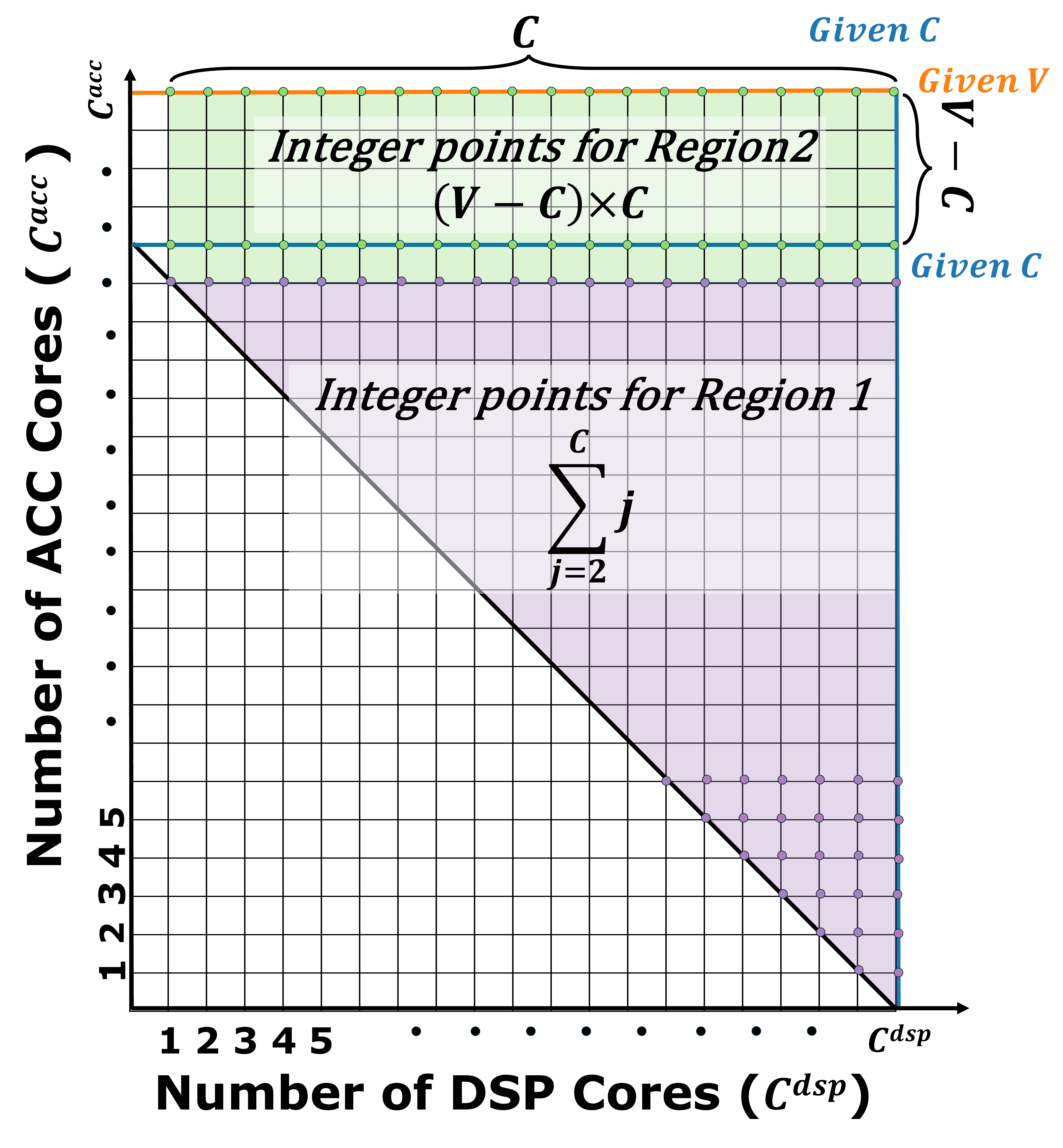}}
    \vspace{-3mm}
    \caption{Region of resource allocations, where integer points in the shaded area represent valid CPU core and VF allocations satisfying {\eqref{eq:core-union}}--{\eqref{eq:core-vf-mapping}}.
    }
    \label{fig:feasible-region}
    \vspace{-5mm}
\end{figure}

\myparatight{5G NR frame structure and numerology.}
The 5G NR frame structure supports multiple numerologies for operation across a wide range of carrier frequencies, including the mmWave bands in FR2. FR2 spans {24.25}\thinspace{GHz}–{52.6}\thinspace{GHz} and commonly utilizes numerology~3 ($\mu=3$), which corresponds to a subcarrier spacing (SCS) of {120}\thinspace{kHz}. This work therefore focuses on $\mu=3$.  
Compared to the standard {15}\thinspace{kHz} subcarrier spacing used in Frequency Range 1 (FR1,  sub-{6}\thinspace{GHz}), the higher SCS of {120}\thinspace{kHz} reduces OFDM symbol duration, making it more suitable for high Doppler spreads and low-latency requirements.
At $\mu=3$, each slot spans {0.125}\thinspace{ms} with 14 symbols, 8$\times$ shorter than the {1}\thinspace{ms} slot at {15}\thinspace{kHz}~\cite{3gpp38.213}. 

\myparatight{Baseband Processing.}
Baseband processing refers to the digital signal processing tasks required to encode and decode user data for wireless transmission. Fig.~\ref{fig:System-Design}(b) shows uplink operations such as FFT, channel estimation, equalization, demodulation, and forward error correction (FEC) before the MAC layer. These tasks are performed after the radio unit (RU) captures the user's signal and streams the raw I/Q samples to the processing unit.
5G NR functional split options~\cite{3gpp38.801} define how these tasks are partitioned across the radio access network (RAN). This work focuses on split option 8, where the entire baseband stack is centralized in the distributed unit (DU). 
This configuration enables flexible compute resource pooling across multiple cells, which is critical in supporting dynamic, multi-cell vRAN deployment.
Notably, uplink baseband processing is significantly more computationally intensive than downlink due to the iterative nature of the brief propagation. As such, {\name} focuses exclusively on uplink processing. 
Prior studies such as FlexRAN~\cite{intel_flexran} and Atlas~\cite{xing2023enabling} report that downlink symbol processing latency is typically $\sim$3$\times$ lower than uplink.  
Thus, demonstrating support for 100\% uplink traffic load in {\name} implicitly ensures that the system can also sustain intensive downlink workloads under the 5G slot format.  
%

\begin{table}[!t]
\centering
\footnotesize
\caption{Parameters of considered \emph{cell complexity}, denoted by $\cellConfigVar(\cellConfigMIMO, \cellConfigBW, \cellConfigTL, \cellConfigTBW, \cellConfigMCS)$ and based on~\cite{3gpp38.214}.}
\label{tab:config-space}
\vspace{-3mm}
\begin{tabular}{llll}
\toprule
\textbf{Dimension} & \textbf{Notation} & \textbf{Range} & \textbf{Granularity} \\
\midrule
Channel BW & $\cellConfigBW$ & \{100, 200, 400\}\thinspace{MHz} & Discrete \\
MIMO Size & $\cellConfigMIMO$ & \{1, 2, 4\} & Discrete \\
Traffic Load & $\cellConfigTL$ & [0,100]\thinspace{\%} & {6.25\%} \\
Transmission BW \% & $\cellConfigTBW$ & [0,100]\thinspace{\%} & Multiple PRBs \\
MCS Index & $\cellConfigMCS$ & $\{0,1, \dots, 28\}$ & Integer \\
\bottomrule
\vspace{-5mm}
\end{tabular}
\end{table}

\section{{\namebf} for a Single Cell}
\label{sec:single-cell}


Even though {\name} is designed to virtualize baseband processing across multiple mmWave cells, the effectiveness of multi-cell scheduling fundamentally depends on accurately understanding how compute resource allocation impacts baseband processing latency for a single cell against varying cell configurations.
Our single-cell profiling and analysis reveal how CPU cores and ACC100 VFs interact, how power and latency trade off across resource allocations, and how cell configuration parameters shape computational demand.  


However, two challenges make single-cell scheduling non-trivial.
First, the possible resource allocation space is large: one cell can map to many combinations of CPU cores and ACC100 VFs, each with different power–latency characteristics. 
Second, feasibility, \emph{defined as whether the 99.9\textsuperscript{th} percentile processing latency under a given allocation meets the strict 3-slot deadline}, is not deterministic but varies non-linearly with different cell configurations.  
To address these challenges, we first model the single-cell allocation space and quantify its brute-force complexity.
We then introduce a random forest (RAF)-based feasibility model that efficiently predicts whether a given allocation can meet real-time processing deadlines, while significantly pruning the search space.
This single-cell framework enables {\name} to select energy-efficient resource allocations, and also serves as the building block for the multi-cell scheduler described in \S\ref{sec:multi-cell}.

\subsection{Cell Configuration}

Let $\cellConfigVar(\cellConfigMIMO, \cellConfigBW, \cellConfigTL, \cellConfigTBW, \cellConfigMCS)$ denote the \emph{cell configuration}, which is a function of the cell's PHY parameters and workloads.
Table~\ref{tab:config-space} summarizes the range and granularity of these parameters based on the 3GPP Technical Specifications for FR2~\cite{3gpp38.214}.
Specifically, we consider the \texttt{DDDSU} TDD frame structure, which allocates 52 downlink, 2 guard, and 16 uplink symbols across five slots, which is widely adopted in both industry and research~\cite{gsma20205g_tdd_sync, lazarev2023resilient, mahapatra2022intel}, and is also recommended by 3GPP for its latency benefits~\cite{fezeu2023mid}.
Since we focus on uplink processing, $\cellConfigTL$ effectively has a granularity of $1/16 = 6.25\%$. 
The transmission BW percentage is defined relative to the maximum number of physical resource blocks (PRBs) supported for a given channel bandwidth.
For example, with a {400}\thinspace{MHz} channel at numerology $\mu=3$, the maximum is 264 PRBs (see Table~5.3.2-1 in~\cite{transmission_BW}). 
This corresponds to a granularity of $1/264 \approx 0.38\%$, where each PRB corresponds to 12 subcarriers, i.e., $12 \times 120\thinspace\text{kHz} = {1.44}\thinspace\text{MHz}$.

\subsection{Optimizing Heterogeneous Compute Resource Allocation}

    

\begin{figure}[!t]
    \centering    
    \subfloat{
    \includegraphics[width=0.80\columnwidth]{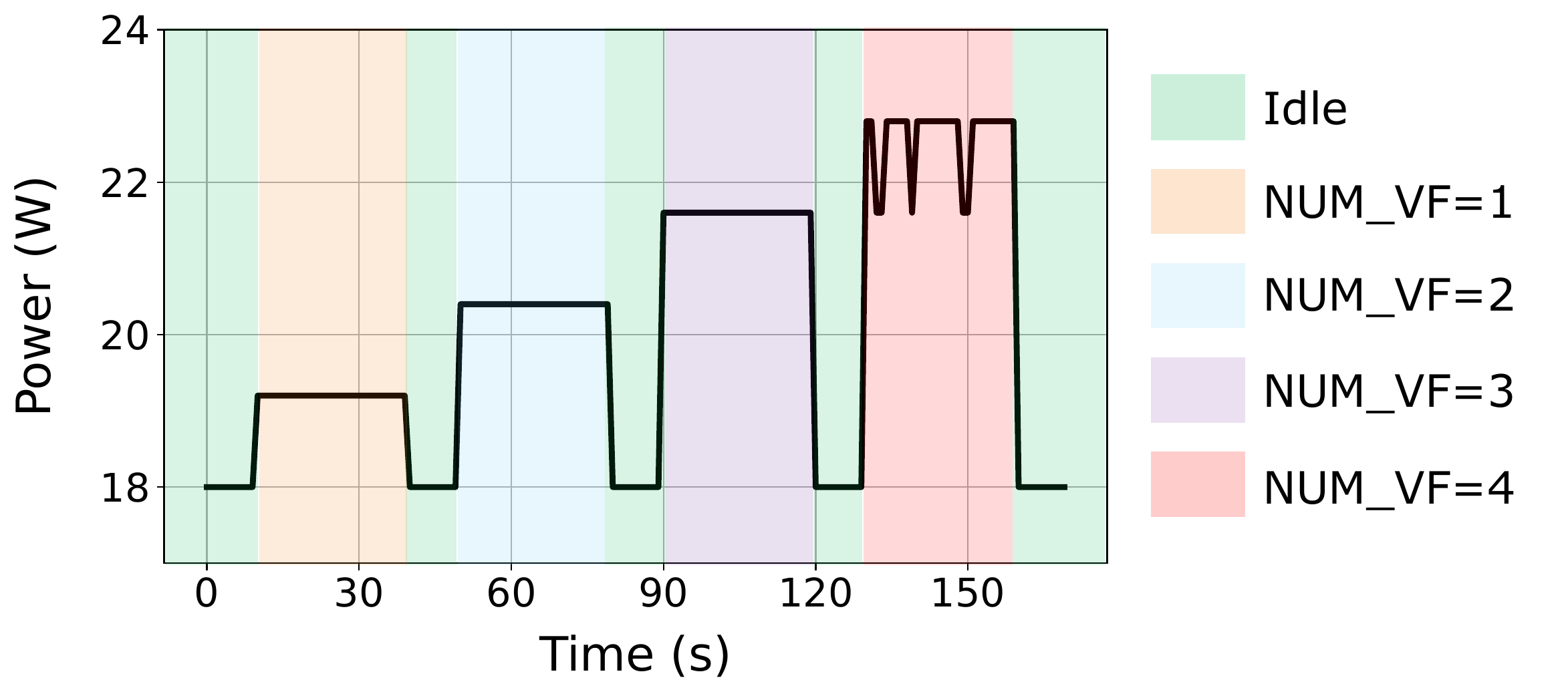}}
    \vspace{-3mm}
    \caption{Power consumption of ACC100 with multiple VFs.}
    \label{fig:VF Power}
    \vspace{-3mm}
\end{figure}

Consider a server equipped with $\numCoreMax$ CPU cores and supporting $\numVfAccMax$ ACC100 accelerator virtual functions (VFs).
We assume $\numCoreMax \geq \numVfAccMax$, e.g., the server used in our experiments is with $\numCoreMax = 56$ and $\numVfAccMax=16$ using a single ACC100 accelerator card.
To support a single cell using a set of CPU cores, $\setCoreTotal$, and a set of ACC100 VFs, $\setVfAcc$, we define the \emph{single-cell resource allocation strategy} as
$\sysConfig{\setCoreTotal}{\setCoreDsp}{\setCoreAcc}{\setVfAcc}$.
Here, $\setCoreDsp$ and $\setCoreAcc$ represent the sets of CPU cores allocated to software DSP and ACC100 offloading tasks, which are optimization variables given the available resources $\setCoreTotal$ and $\setVfAcc$.
Since all CPU cores are considered homogeneous and equivalent, the specific sets of cores assigned to different tasks are irrelevant; only the number of cores allocated to each task matters.
Therefore, without loss of generality, this single-cell resource allocation can be represented by $\sysConfig{\numCoreTotal}{\numCoreDsp}{\numCoreAcc}{\numVfAcc}$, where $\numCoreDsp = \setSize{\setCoreDsp}$, $\numCoreAcc = \setSize{\setCoreAcc}$, $\numCoreTotal = \setSize{\setCoreTotal}$, and $\numVfAcc = \setSize{\setVfAcc}$.

\myparatight{Feasible resource allocation strategies.}
For a given number of $\numCoreTotal \leq \numCoreMax$ and $\numVfAcc \leq \numVfAccMax$, the heterogeneous compute resource allocation for a single cell needs to satisfy the following constraints describing the relationship between the DSP pipeline and utilized hardware resources:
\begin{itemize}[leftmargin=*, topsep=2pt, itemsep=1pt]
\item
Each CPU core is dedicated to either software DSP, offloading LDPC decoding to ACC100, or both, i.e.,
\begin{align} 
    1 & \leq \numCoreTotal = \setSize{\setCoreTotal} = \setSize{\setCoreDsp \cup \setCoreAcc} \leq \numCoreDsp + \numCoreAcc, \notag \\
    1 & \leq \numCoreDsp \leq \numCoreTotal~\textrm{and}~
    1 \leq \numCoreAcc \leq \numCoreTotal.
    \label{eq:core-union} 
\end{align}
When ACC100 VFs are available, LDPC decoding is offloaded to the hardware accelerator, while the CPU cores are responsible for all remaining DSP tasks.
\item
Each CPU core responsible for ACC100 offloading tasks is mapped to a unique set of VFs, and each VF cannot be managed by more than one core, i.e.,
\begin{align} 
    1 \leq \numCoreAcc \leq \numVfAcc.
    \label{eq:core-vf-mapping} 
\end{align}
\end{itemize}
For fixed values of $\numCoreTotal$ ($\leq \numCoreMax$) and $\numVfAcc$ ($\leq \numVfAccMax$), let $\setSysConfig{\numCoreTotal}{\numVfAcc}$ denote the set of feasible allocations satisfying the abovementioned constraints, i.e.,
\begin{align}
    \setSysConfig{\numCoreTotal}{\numVfAcc} = \left\{ \sysConfig{\numCoreTotal}{\numCoreDsp}{\numCoreAcc}{\numVfAcc}:~\textrm{\eqref{eq:core-union}--\eqref{eq:core-vf-mapping} are satisfied} \right\}.
    \label{eq:set-config-given-c-v}
\end{align}
Fig.~\ref{fig:feasible-region} illustrates this feasibility region (shaded areas), with each integer point $(\numCoreDsp, \numCoreAcc)$ representing one feasible resource allocation strategy.
It can be shown that the number of feasible allocations is:
\begin{align}
    \setSize{\setSysConfig{\numCoreTotal}{\numVfAcc}}
    =
    \begin{cases}
        \sum_{j=1}^{\numVfAcc} (j+1) = \frac{\numVfAcc(\numVfAcc+3)}{2},\ & \textrm{if}~\numVfAcc \leq \numCoreTotal, \\
        \sum_{j=2}^{\numCoreTotal} j + (\numVfAcc-\numCoreTotal)\cdot\numCoreTotal = \frac{2\numVfAcc\numCoreTotal + \numCoreTotal -\numCoreTotal^2 -2}{2},\ & \textrm{if}~\numVfAcc > \numCoreTotal.
    \end{cases}
\end{align}

\myparatight{Power-aware resource optimization.}
{\name} addresses the challenge of efficiently selecting the allocation strategy to meet the stringent real-time PHY processing deadline in a virtualized baseband system, while minimizing the power consumption.
This power-aware resource optimization is
\begin{align}
\hspace{-5mm}
\sysConfigOpt := \arg\min_{\sysConfigVar}:~
& \power(\sysConfigVar) = \alpha_{1} P_{1}(\numCoreDsp) + \alpha_{2} P_{2}(\numCoreAcc) + \alpha_{3} P_{3}(\numVfAcc)\,
\label{eq:single-cell-opt-obj-full} \\
\textrm{s.t.}:~ & \latency(\sysConfigVar) < \textrm{3-slot},\ \sysConfigVar \in \setSysConfig{\numCoreTotal}{\numVfAcc}~\textrm{from}~{\eqref{eq:set-config-given-c-v}}, \nonumber \\
& 1 \leq \numCoreTotal \leq \numCoreMax,\ 1 \leq \numVfAcc \leq \numVfAccMax. \nonumber
\end{align}
Here, $P_{1}(\numCoreDsp)$, $P_{2}(\numCoreAcc)$, and $P_{3}(\numVfAcc)$ denote the power consumption of DSP worker cores, ACC100 offloading cores, and ACC100 VFs, respectively.
The weights $\{\alpha_{1}, \alpha_{2}, \alpha_{3}\}$ allow system designers to bias resource selection depending on hardware constraints, which are set to equal in {\name}, i.e., $\alpha_{1}=\alpha_{2}=\alpha_{3}=1$).
%
%
We use Intel’s Performance Counter Monitor (PCM)~\cite{intelpcm} to measure the CPU power consumption.
Each active core contributes approximately {7}\thinspace{W} of dynamic power, consistent with prior studies~\cite{qi2024savannah}, yielding
$P_{1}(\numCoreDsp) + P_{2}(\numCoreAcc) = 7 \cdot \numCoreTotal.$
To characterize the power consumption of the ACC100, we conduct measurements on an ACC100 connected to a workstation through a PCIe riser cable and isolate the {12}\thinspace{V} supply for inline current measurement using a clamp meter following prior work~\cite{qi2024savannah, romein2018powersensor}.
The measurement accuracy of this approach is validated using an NVIDIA GTX1070 GPU with the \texttt{\small nvidia-smi} tool.
As shown in Fig.~\ref{fig:VF Power}, each active ACC100 VF adds approximately {1.2}\thinspace{W} power, yielding
$P_{3}(\numVfAcc) = 1.2 \cdot \numVfAcc.$
%

\subsection{Profiling and Exploring Resource Allocation Strategies}
\label{ssec:single-cell-opt}



\myparatight{Profiling methodology.}
To understand how resource allocations $\sysConfigVar$ impact feasibility, we developed and instrumented the {\name} software stack to record per-frame processing latency across diverse cell configurations.  
Built on top of the Agora~\cite{ding2020agora} and {\savannah}~\cite{qi2024savannah} codebases, {\name} extends support for multiple implementations, each corresponding to a distinct resource allocation strategy.  
This required substantial engineering effort to enable flexible switching between software-only, hardware-accelerated, and hybrid paths (\S\ref{sec:implementation}).  
%
%
For each $\cellConfigVar$ under various $\sysConfigVar$, we execute 20K frames and record the 99.9th percentile latency.
Due to space limit, Fig.~\ref{fig:MOO} reports representative power–latency trade-offs with varying $(\cellConfigMIMO, \cellConfigBW)$ and $\cellConfigTL \in \{25\%, 100\%\}$ under $MCS=17$.
Similar trends are observed when varying other cell configuration parameters such as $\cellConfigTBW$ or $\cellConfigMCS$.
From these measurements, we derive five key observations that guide the design of {\name}.  
%
%

\myparatight{Observation 1: The mapping from cell configuration $\cellConfigVar$ to feasibility is highly non-linear.}
Our measurements show that whether a cell meets the 3-slot deadline cannot be explained by a single feature.
Instead, feasibility depends on complex cross-terms—e.g., a 2$\times$2 {100}\thinspace{MHz} cell with $\cellConfigTL=100\%$ may exceed the deadline under a given allocation, while a 4$\times$4 {100}\thinspace{MHz} cell at lower load may remain feasible. 
This non-straightforward mapping motivates the need for systematic profiling and learning–based models.

\myparatight{Observation 2: A \emph{low-complexity} cell can be fully handled in software by a single CPU core without requiring accelerator offloading, e.g., $\sysConfig{1}{1}{0}{0}$.}
Under light traffic, the baseband workload processed by a single worker core can meet the deadline without offloading LDPC decoding to the accelerator.
For example, a single CPU core can support a 1$\times$1 {400}\thinspace{MHz} cell with $\cellConfigTL = 25\%$ (Fig.~\ref{figs:MOO--SISO-400MHz-25TL}). 

\myparatight{Observation 3: A \emph{low-complexity} cell requiring additional decoding capacity can benefit from adding a single VF alongside one DSP core, e.g., $\sysConfig{1}{1}{1}{1}$.}
Evaluating and building on the prior work {\savannah}~\cite{qi2024savannah}, $\sysConfig{1}{1}{1}{1}$ can sustain a full traffic load ($\cellConfigTL=100\%$) for a 1$\times$1 {100}\thinspace{MHz}, 1$\times$1 {200}\thinspace{MHz}, or 2$\times$2 {100}\thinspace{MHz} cell.
However, this resource allocation is inadequate with increased cell complexity.
For instance, the same allocation can only sustain a traffic load of $\cellConfigTL=50\%$ for a 1$\times$1 {400}\thinspace{MHz} cell, $\cellConfigTL=44\%$ for 2$\times$2 {200}\thinspace{MHz}, and $\cellConfigTL=38\%$ for 4$\times$4 {100}\thinspace{MHz}.
The bottleneck arises from software DSP stages saturating the single CPU core and from the rising number of MIMO streams feeding the decoder.

\myparatight{Observation 4: A \emph{moderate-complexity} cell can be supported by multiple DSP cores and one VF, e.g., $\sysConfig{c}{c-1}{1}{1}$ or $\sysConfig{c}{c}{1}{1}$.}
Supporting moderate-complexity cells requires allocating $c>1$ CPU cores to software DSP while reserving a single VF for LDPC decoding.
Our profiling reveals that the effectiveness of $\sysConfig{c}{c-1}{1}{1}$ versus $\sysConfig{c}{c}{1}{1}$ depends on the number of available cores.
When $\numCoreTotal$ is small (e.g., two or three), the bottleneck lies in software DSP.
In such cases, using all cores for DSP, as in $\sysConfig{2}{2}{1}{1}$ for a 2$\times$2 {200}\thinspace{MHz} cell, outperforms $\sysConfig{2}{1}{1}{1}$ and meets the deadline without requiring the extra core that $\sysConfig{3}{2}{1}{1}$ would consume. 
However, as $\numCoreTotal$ increases, DSP no longer becomes the limiting factor.
Here, isolating one core dedicated to ACC100 offloading, i.e., $\sysConfig{c}{c-1}{1}{1}$, effectively eliminates queuing delays at the decoder for a more stable system.

\myparatight{Observation 5: A \emph{high-complexity} cell (e.g., larger $\cellConfigMIMO$, wider $\cellConfigBW$, higher $\cellConfigMCS$) requires full offloading across multiple cores and VFs, e.g., $\sysConfig{c}{c}{c}{c}$.}
For high-complexity cells (e.g., 4$\times$4 {100}\thinspace{MHz}), increasing the number of DSP cores alone cannot close the latency gap because a single VF becomes the bottleneck: each additional MIMO stream produces proportionally more code blocks, saturating the decoder.
While decoding a 1$\times$1 {100}\thinspace{MHz} cell consumes only about {0.38}\thinspace{Gbps} ($\approx$5\% of ACC100's peak throughput), a 4$\times$4 cell scales this demand beyond a single VF’s capacity.
To address this, we partition the ACC100 into multiple VFs, distributing decoding tasks in parallel. 
Balanced configurations in the form of $\sysConfig{c}{c}{c}{c}, \forall c = 1, \dots, \min\{C,V\}$m
ensure that each DSP core is ``paired'' with its own VF, avoiding queueing imbalance.
Across diverse cells, this balanced resource allocation strategy achieves the highest throughput, while for low-complexity cells, $\sysConfig{c}{c-1}{1}{1}$ may still remain preferable due to reduced VF usage and thus power consumption.

\myparatight{Reduced resource allocation space for a single cell.}
The above observations, validated through extensive experiments, allow us to constrain the search to a small but representative subset. 
In addition, a clear flattening effect can be observed from Fig.~\ref{fig:MOO}: as more CPU cores are added, the processing latency eventually stops improving.
Beyond this point, allocating additional cores yields diminishing returns. 
Guided by these observations, we define the reduced feasible resource allocation space for a single cell as:
\begin{equation}
\label{eq:set-config-given-c-v-reduced}
\begin{aligned}
    \setSysConfigReduced{\numCoreTotal}{\numVfAcc}
    = \{ 
    & \sysConfig{c}{c}{0}{0}, && c \in \{1,2,\cdots,6\}, \\
    & \sysConfig{c}{c}{1}{1}, && c \in \{1,2,3\}, \\
    & \sysConfig{c}{c-1}{1}{1}, && c \in \{4,5,6\}, \\
    & \sysConfig{c}{c}{c}{c}, && c \in \{3,4,5\}
    \}.
\end{aligned}
\end{equation}

Unlike the full allocation space $\setSysConfig{\numCoreTotal}{\numVfAcc}$ in {\eqref{eq:set-config-given-c-v}}, which grows combinatorially, the reduced set has a fixed cardinality of $\setSize{\setSysConfigReduced{\numCoreTotal}{\numVfAcc}} = 15$ independent of $\numCoreTotal$ and $\numVfAcc$ as long as $\min\{\numCoreTotal, \numVfAcc\} \geq 6$, which is common for a many-core server equipped with a single ACC100 accelerator.
For example, with $\numCoreTotal=32$ and $\numVfAcc=16$, the reduced set still contains only $15$ configurations, compared to $152$ in the original space.


\begin{figure*}[!t]
    \centering
    \includegraphics[width=0.65\textwidth]{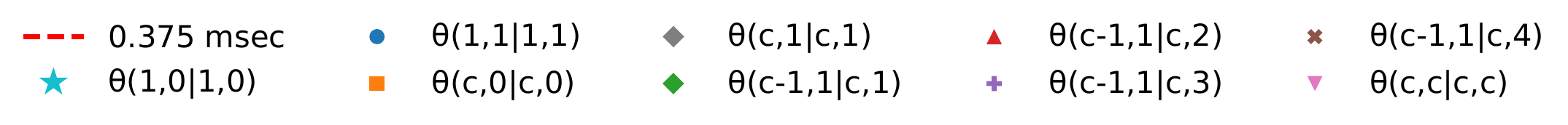}
    \\
    \vspace{-4mm}
    \subfloat[\hspace{4mm}1$\times$1 {400}\thinspace{MHz} - 25\% $\cellConfigTL$]{
    \label{figs:MOO--SISO-400MHz-25TL}
    \includegraphics[width=0.28\textwidth]{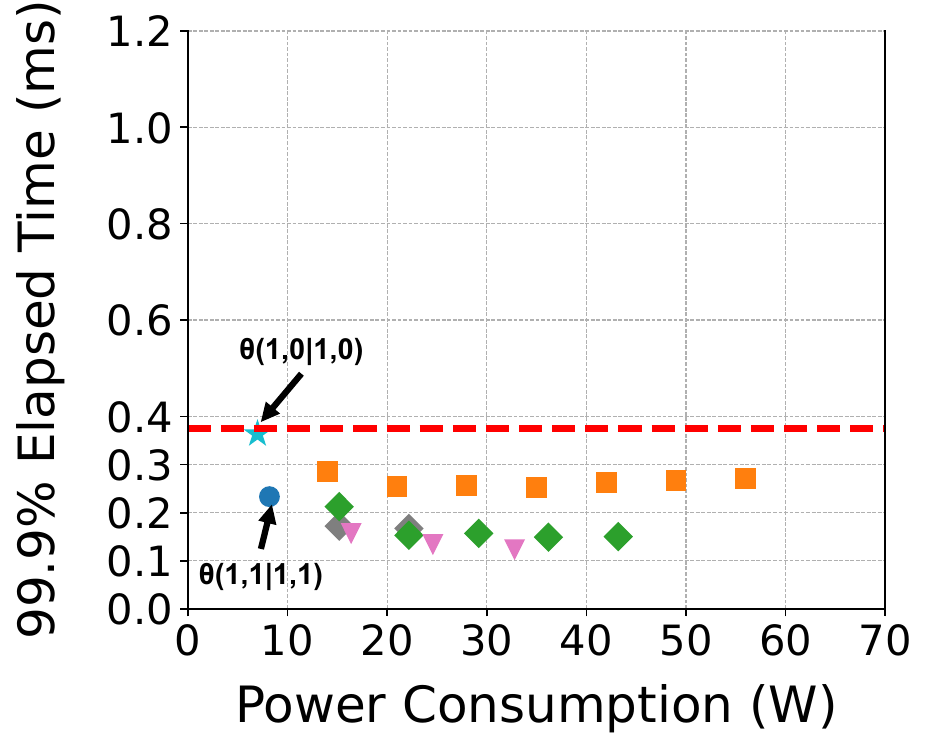}}
    \hspace{3mm}
    \subfloat[\hspace{4mm} 2$\times$2 {200}\thinspace{MHz} - 25\% $\cellConfigTL$]{
    \label{figs:MOO--2-2-200MHz-25TL}
    \includegraphics[width=0.28\textwidth]{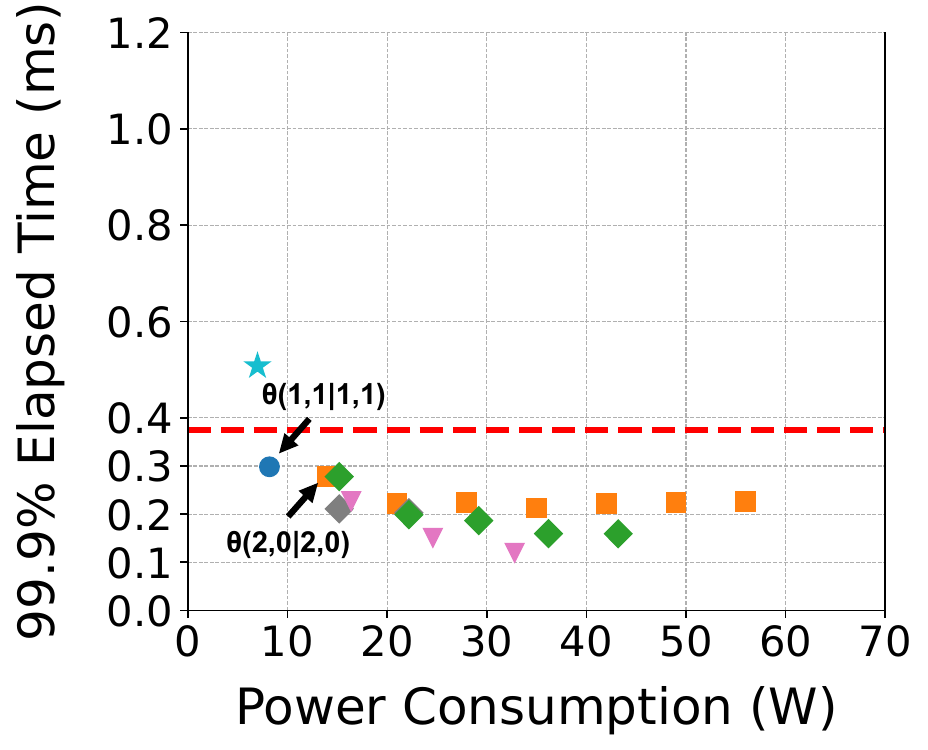}}
    \hspace{3mm}
    \subfloat[\hspace{4mm} 4$\times$4 {100}\thinspace{MHz} - 25\% $\cellConfigTL$]{
    \label{figs:MOO--4-4-100MHz-25TL}
    \includegraphics[width=0.28\textwidth]{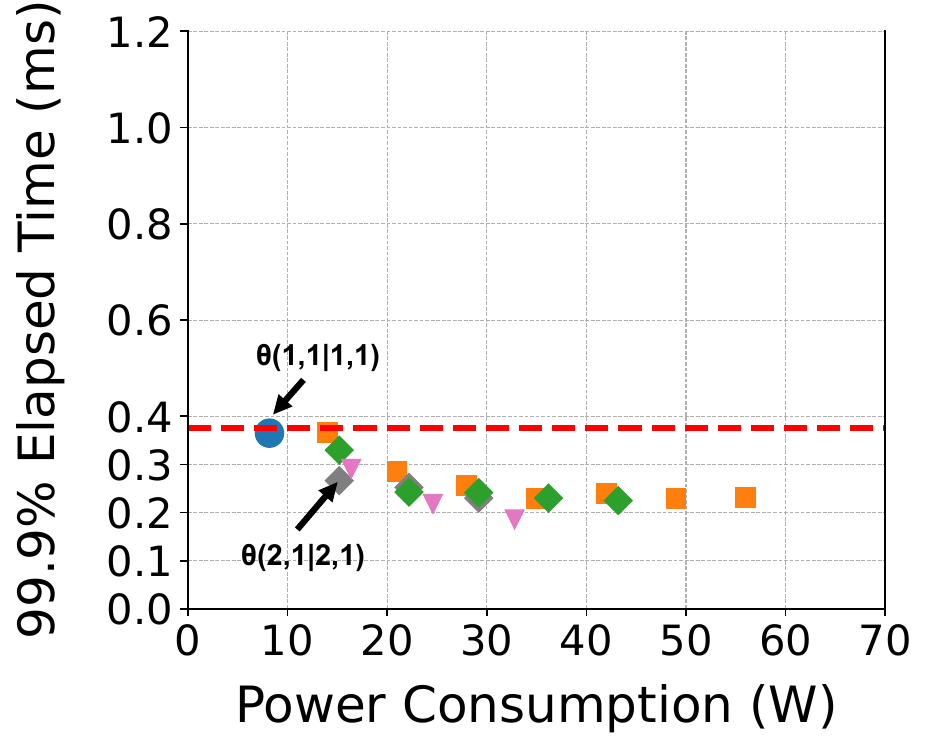}}
    \\

    \subfloat[\hspace{4mm}1$\times$1 {400}\thinspace{MHz} - 100\% $\cellConfigTL$]{
    \label{figs:MOO--SISO-400MHz}
    \includegraphics[width=0.28\textwidth]{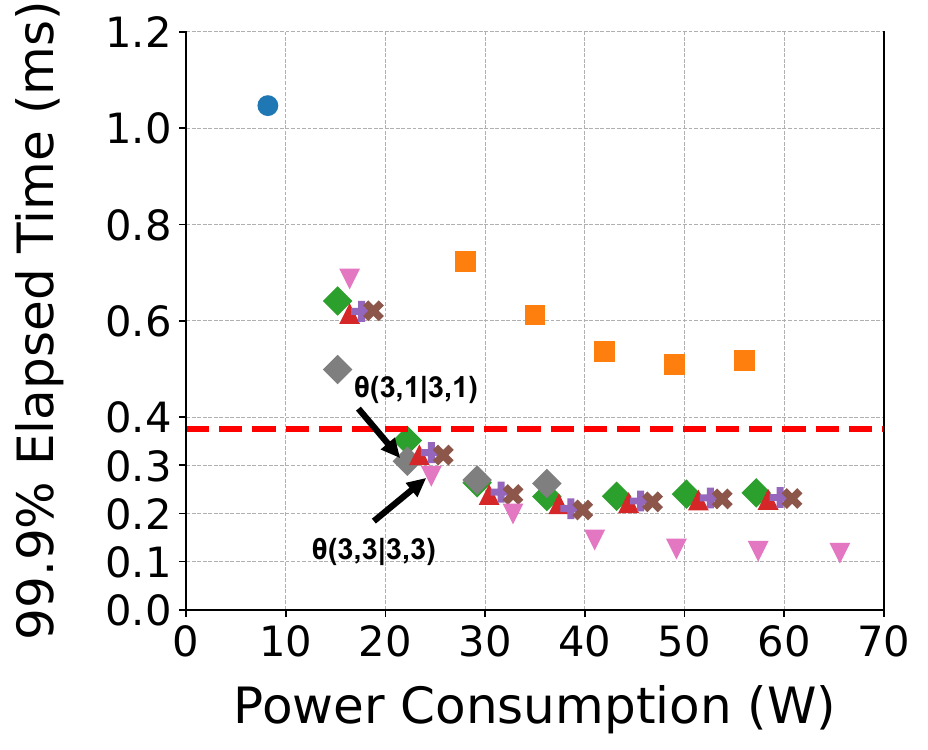}}
    \hspace{3mm}
    \subfloat[\hspace{4mm} 2$\times$2 {200}\thinspace{MHz} - 100\% $\cellConfigTL$]{
    \label{figs:MOO--2-2-200MHz}
    \includegraphics[width=0.28\textwidth]{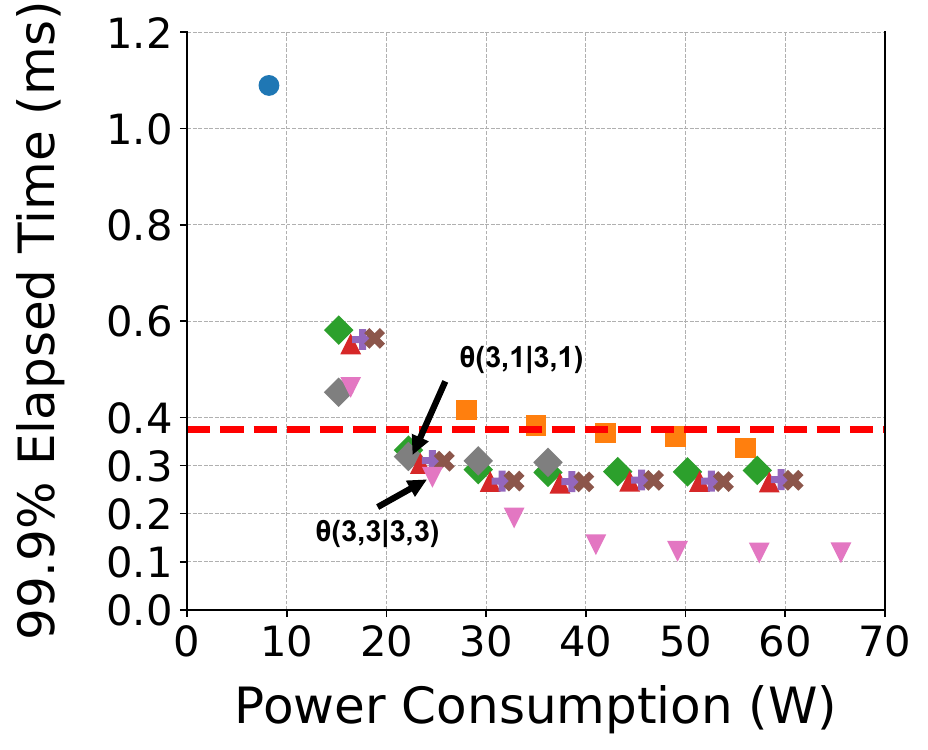}}
    \hspace{3mm}
    \subfloat[\hspace{4mm} 4$\times$4 {100}\thinspace{MHz} - 100\% $\cellConfigTL$]{
    \label{figs:MOO--4-4-100MHz}
    \includegraphics[width=0.28\textwidth]{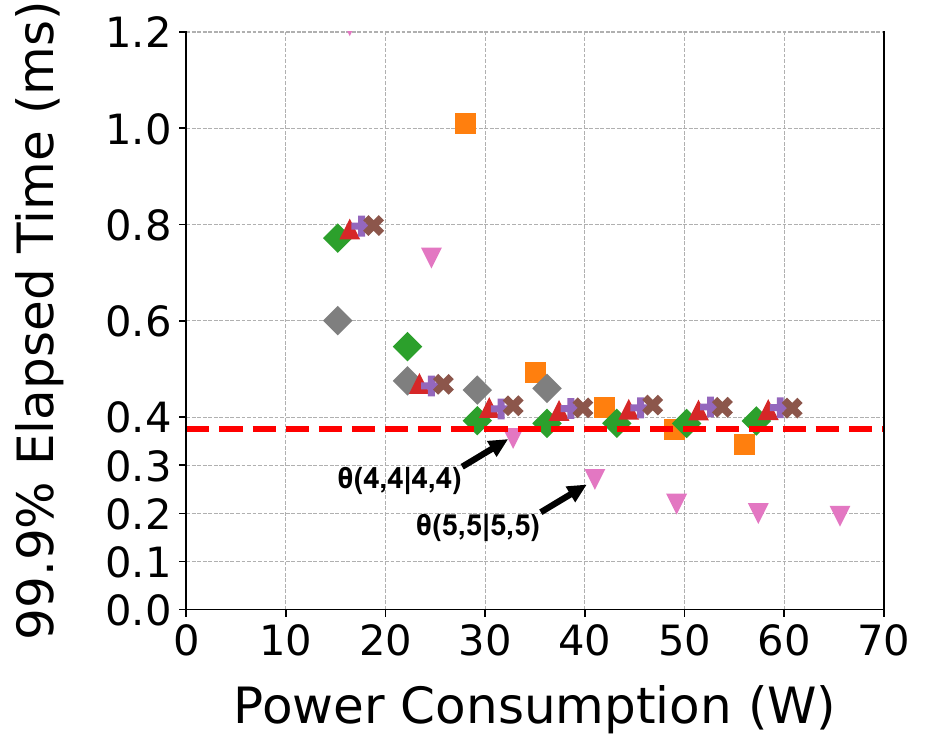}}
    \vspace{-3mm}
    \caption{Power-latency trade-off space for {\namebf} across heterogeneous cell configurations.
    Each point represents a distinct resource allocation. Labeled dots indicate the most and second most energy-efficient options that {\namebf} is likely to select.
    }
    \label{fig:MOO}
    \vspace{-3mm}
\end{figure*}


\subsection{{\namebf}'s Single-Cell Scheduler}
\label{ssec:RAF-model}
While the power model in {\eqref{eq:single-cell-opt-obj-full}} quantifies the cost of a given resource allocation, it does not establish whether that allocation can meet the 3-slot deadline.  
Simple rule-based pruning is insufficient: while it captures broad regimes 
it cannot accurately characterize intermediate cases where feasibility hinges on subtle cross-terms
and CPU-ACC100 coupling effects.  
This motivates the need for a supervised learning model that can generalize from profiling data and reliably map $(\cellConfigVar,\sysConfigVar)$ to feasibility.  

To overcome this challenge, {\name} augments profiling with a supervised learning model that generalizes feasibility from sampled data.  
Specifically, we train a random forest (RAF) classifier that, for each $(\cellConfigVar,\sysConfigVar)$ pair, outputs a confidence score
$f(\cellConfigVar,\sysConfigVar) \in [0,1]$,
representing the probability that the allocation $\sysConfigVar$ can meet the 3-slot deadline under configuration $\cellConfigVar$ (see \S\ref{sec:evaluation} for details).  
We then define feasibility relative to a decision threshold $\tau$ (e.g., $\tau=0.5$): if $f(\cellConfigVar,\sysConfigVar) \geq \tau$, the allocation is labeled \emph{feasible}; otherwise it is \emph{infeasible}.  
At runtime, given a new and potentially unseen $\cellConfigVar$, {\name}'s single-cell scheduler evaluates each candidate $\sysConfigVar$ from the reduced search space and selects the feasible subset:
\begin{align}
\cellFeasibleRF{\cellConfigVar} 
= \Big\{ \sysConfigVar \in \setSysConfigReduced{\numCoreTotal}{\numVfAcc} \;\Big|\; f(\cellConfigVar,\sysConfigVar) \geq \tau \Big\}.
\end{align}


Replacing the full allocation space with the reduced search set given by\eqref{eq:set-config-given-c-v-reduced}, {\name}'s RFA-based single-cell scheduler is:
\begin{tcolorbox}[boxsep=0mm,left=1mm,right=1mm,top=-1mm,bottom=1mm,lefttitle=2mm,
 title={Optimizing single-cell resource allocation \textsf{(Opt-$1$-Cell)}},
 boxrule=1pt,sharp corners=all]
\begin{align}
\sysConfigOpt := & \arg\min_{\sysConfigVar \in \setSysConfigReduced{\numCoreTotal}{\numVfAcc}} 
      \;\power(\sysConfigVar) = 7 \cdot \numCoreTotal + 1.2 \cdot \numVfAcc \nonumber \\
\text{s.t.}\quad & f(\cellConfigVar,\sysConfigVar)\geq \tau \nonumber.
\label{eq:single-cell-opt-obj}
\end{align}
\end{tcolorbox}

\noindent Solving \textsf{(Opt-1-Cell)} returns the most energy-efficient resource allocation $\sysConfigOpt$, i.e., the number of CPU cores and ACC100 VFs that the cell should use while still meeting the deadline.  
For example, Fig.~\ref{figs:MOO--SISO-400MHz} shows that while multiple allocations are feasible, \textsf{(Opt-1-Cell)} selects $\sysConfig{3}{3}{1}{1}$ for a 1$\times$1 {400}\thinspace{MHz} cell, achieving the lowest power of {22.2}\thinspace{W}.  
At runtime, the scheduler instantiates this decision by assigning the corresponding physical worker cores and VFs to the cell’s processing pipeline (see \S\ref{sec:implementation} for details).  

\myparatight{Model selection rationale.}
We select RAF as the feasibility predictor for two reasons. 
(\emph{i}) \emph{Capturing non-linear interactions:} feasibility boundaries depend on cross-terms such as $\cellConfigTL$, $\cellConfigBW$, $\cellConfigMIMO$, etc. 
More importantly, interactions between CPU and ACC100 offloading introduce additional variability (e.g., bbdev queueing, PCIe transfer) that are extremely difficult to model analytically.
(\emph{ii}) \emph{Data efficiency and speed:}
compared with DL models, RAF achieves competitive accuracy with substantially less training data and provides real-time $\mu$s-level inference per $(\cellConfigVar, \sysConfigVar)$.
As shown in \S\ref{sec:evaluation}, our evaluation against six baselines, 
demonstrates that RAF delivers the best balance of accuracy, inference latency, and model size. 

\section{{\namebf} for Multiple Cells}
\label{sec:multi-cell}

Based on the single-cell resource allocation analysis, we now extend {\name} to support concurrent multi-cell baseband processing with scalability.
{\name}, as illustrated in Fig.~\ref{fig:System-Design}, achieves this through two key components:
(\emph{i}) the parallel instantiation of multiple single-cell processing pipelines, and
(\emph{ii}) a power-aware scheduler that allocates heterogeneous compute resources while considering resource contention.

\subsection{Multi-cell Resource Allocation Model}

While the formulation $\sysConfig{\numCoreTotal}{\numCoreDsp}{\numCoreAcc}{\numVfAcc}$ in \S\ref{sec:single-cell} focuses on single-cell configuration, {\name} is designed to support multi-cell deployments. Consider the scenario where $N$ cells need to be served by a pool of worker cores and ACC100 VFs.
The $i$-th cell ($i = 1, 2, \dots, N$) with cell configuration $\cellConfigVar_i$,
may adopt a different resource allocation strategy, denoted by $\sysConfigCellIdx{i}{\numCoreTotalCellIdx{i}}{\numCoreDspCellIdx{i}}{\numCoreAccCellIdx{i}}{\numVfAccCellIdx{i}}$.
In this context, we assume that all compute resources, worker cores, and VFs, are assigned in a non-overlapping manner across the cells, i.e., 
\begin{equation}
\label{eq:nonoverlap}
\setCoreTotal_{i} \cap \setCoreTotal_{j} = \varnothing ~\textrm{and}~ \setVfAcc_{i} \cap \setVfAcc_{j} = \varnothing,~ \forall 1 \leq i \neq j \leq N.
\end{equation}
This assumption on non-overlapping resource allocation is both practical and widely adopted. 
Academic platforms (e.g., Agora~\cite{ding2020agora}, Hydra~\cite{gong2023scalable}) and industry-grade implementations (e.g., srsRAN~\cite{srsran}, OAI~\cite{OAI}, Intel FlexRAN~\cite{intel_flexran}) all dedicate physical cores to baseband tasks by isolating cores, disabling hyper-threading, and using real-time priorities. 
Sharing cores across cells introduces non-deterministic context-switching delays: for example, on modern Intel Xeon CPUs, a single round of context switches among four cells can incur $\approx$60$\mu$s, which is already 20\% of the 3-slot deadline. 

To enable concurrent allocations, {\name} adopts a modular architecture with multiple processing pipelines, each bound to a resource allocation $\sysConfigVar_{i} \in \setSysConfigReduced{\numCoreTotal_{i}}{\numVfAcc_{i}}$ for cell $i$. Pipelines are statically compiled but dynamically selected at runtime. At execution, the scheduler determines the appropriate $\sysConfigVar_{i}$ for each active cell and injects the corresponding resource allocation strategy as a configuration entry,
activating the desired data path
without recompilation.
The modularity allows {\name} to support heterogeneous multi-cell deployments
with scalability and extensibility: {\name} can easily incorporate future accelerators (e.g., GPUs) or hybrid processing models to further maximize resource utilization and system performance.

\subsection{{\namebf} Multi-Cell Scheduler}
\label{ssec:scheduler}

Given a pool of $\numCoreMax$ worker cores and $\numVfAccMax$ ACC100 VFs that can be shared to concurrently serve $N$ cells, {\name} aims to minimize the overall power consumption when allocating compute resources to individual cells, $N$ cells to achieve minimized energy, i.e.,
\begin{tcolorbox}[boxsep=0mm,left=1mm,right=1mm,top=-1mm,bottom=1mm,
 toptitle=1mm,bottomtitle=1mm,lefttitle=2mm,
 title={Optimizing multi-cell resource allocation \textsf{(Opt-$N$-Cell)}},
 boxrule=1pt,sharp corners=all]
\begin{align}
\{ \sysConfigOpt_{i} \}_{i=1}^{N} := & \arg\min_{\{\numCoreTotal_{i}\}, \{\numVfAcc_{i}\}}~ \sum_{i=1}^{N} \left( 7 \cdot \numCoreTotal_{i} + 1.2 \cdot \numVfAcc_{i} \right), \nonumber \\
\text{s.t.}\quad & f(\cellConfigVar_{i},\sysConfigVar_{i})\geq \tau, \quad \forall i = 1,\dots,N, \nonumber \\
& \sysConfigVar_{i} \in \setSysConfigReduced{\numCoreTotal_{i}}{\numVfAcc_{i}}~\textrm{from}~{\eqref{eq:set-config-given-c-v-reduced}}, \nonumber\\
& 1 \leq \sum_{i=1}^{N} \numCoreTotal_{i} \leq \numCoreMax,\quad 
  1 \leq \sum_{i=1}^{N} \numVfAcc_{i} \leq \numVfAccMax.
\label{eq:multi-cell-opt-core-vf-count}
\end{align}
\end{tcolorbox}
\noindent
However, \textsf{(Opt-$N$-Cell)} cannot be easily decomposed into $N$ independent single-cell resource allocation problems of the form \textsf{(Opt-$1$-Cell)} for two key reasons.
First, the latency of a given cell allocation is no longer determined solely by its local configuration; it is also influenced by system-wide contention such as shared memory bandwidth and PCIe utilization when multiple pipelines execute concurrently. 
Second, 
in the multi-cell case, these limited resources, $\numCoreMax$ cores and $\numVfAccMax$ VFs, must be partitioned among the $N$ cells under the constraints described in {\eqref{eq:multi-cell-opt-core-vf-count}}. 
%
This yields a combinatorial search space with $\binom{\numCoreMax-1}{N-1} \cdot \binom{\numVfAccMax-1}{N-1}$ possible partitions.
%
To address these challenges, we develop an energy-aware resource scheduler (Algorithm~\ref{alg:profiling_scheduler}).

\noindent
\textbf{Contention-aware feasibility adjustment via RAF confidence.}
Let $p = f(\cellConfigVar_i,\sysConfigVar_i^{(k)}) \in [0,1]$ denotes the RAF confidence score, which reflects how likely a configuration $\sysConfigVar_{i}^{(k)}$ in $\cellFeasibleRF{\cellConfigVar_{i}}$ for cell $i$ satisfies the 3-slot deadline in the single-cell setting. 
Through controlled experiments, we observed a strong correlation between $p$ and the empirical latency margin, i.e., the distance between the measured 99.9-th percentile latency and the 3-slot boundary. 
Hence, we use $p$ as a proxy for how close the system is to violating the deadline: high $p$ indicates substantial margin, while low $p$ indicates the latency is near the boundary.

\noindent
\textbf{Contention-aware latency adjustment via learned linear model.}
To capture the impact of running multiple cells concurrently, we introduce a lightweight linear adjustment model that degrades the effective RAF confidence as the number of active cells $N$ increases. 
Specifically, for cell $i$ and allocation $\sysConfigVar_{i}^{(k)}$, we define the adjusted confidence as
\begin{equation}
\label{eq:rf_confidence_adjusted}
\hat{p}_{i}^{(k)}(N) \;=\; p_{i}^{(k)} - (\beta_0 + \beta_1 \cdot (N-1)),
\end{equation}
where $p_{i}^{(k)}$ is the RAF confidence in the single-cell case, and $(\beta_0,\beta_1)$ are fitted from multi-cell experiments measuring how confidence scores shift as additional cells are activated (example seen in Fig.~\ref{fig: Same-Cell-MCS10}). 
A configuration $\sysConfigVar_{i}^{(k)}$ is then considered feasible under $N$ cells if $\hat{p}_{i}^{(k)}(N)$ remains above a threshold $\tau$. 
This approach allows our system to combine {\name} single-cell scheduling with a contention-aware adjustment that generalizes to multi-cell deployments.

\noindent
\textbf{Greedy initialization guided by RAF feasibility (single-cell oracle).}
For each active cell $i$ with features $\cellConfigVar_i$, we first enumerate the reduced candidate set 
$\setSysConfigReduced{\numCoreTotal_i}{\numVfAcc_i}$ (15 possible allocations), among which we retain only the RAF-feasible subset
\begin{equation}
\label{eq:feasible_rf}
\cellFeasibleRF{\cellConfigVar_i} \;=\; \big\{ \sysConfigVar^{(k)} \in 
\setSysConfigReduced{\numCoreTotal_i}{\numVfAcc_i} \;\big|\; 
f(\cellConfigVar_i,\sysConfigVar^{(k)}) \geq \tau \big\},
\end{equation}
where $f(\cellConfigVar_i,\sysConfigVar^{(k)})=1$ indicates that the RAF classifier predicts the $99.9$-th percentile latency under 
$(\cellConfigVar_i,\sysConfigVar^{(k)})$ meets the 3-slot deadline in the single-cell setting.  

We then order this feasible set by increasing power consumption to form
\begin{align}
\cellProfileRF{\cellConfigVar_i} = \Big\{ 
\sysConfigVar^{(1)}, \dots, & \sysConfigVar^{(K)} ~\Big|~ \sysConfigVar^{(k)} \in \cellFeasibleRF{\cellConfigVar_i}, \nonumber \\
& \power(\sysConfigVar^{(1)}) < \dots < \power(\sysConfigVar^{(K)})
\Big\}.
\label{eq:profile_table_rf}
\end{align}
At runtime, we select the first $\sysConfigVar^{(k)} \in \cellProfileRF{\cellConfigVar_i}$ whose contention-adjusted feasibility
$\hat{p}_{i}^{(k)}(N)$ from~\eqref{eq:rf_confidence_adjusted} satisfies $\hat{p}_{i}^{(k)}(N) \ge \tau$,
indicating the 99.9-th percentile latency is predicted to meet the 3-slot deadline under $N$-cell contention.

\myparatight{Reduced resource allocation space for $N$ cells.}
To support $N$ cells on a system equipped with $\numCoreMax$ CPU cores and $\numVfAccMax$ ACC100 VFs, a brute-force approach would require evaluating all possible resource allocation combinations across cells. The total number of such allocations is given by:
\begin{align}
    \setSize{\{ \bm{\uptheta}_{i} \}_{i=1}^{N}}
    =
    \binom{\numCoreMax - 1}{N - 1}
    \cdot
    \binom{\numVfAccMax - 1}{N - 1}
\end{align}
%
%
%

{\name} addresses this challenge by leveraging its resource allocation scheduler, which reduces the per-cell allocation space to a small fixed set of 15 strategies (Eq.~\ref{eq:profile_table_rf}). As a result, the multi-cell search complexity becomes linear in $N$:
\begin{align}
\setSize{\{ \bm{\uptheta}'_{i} \}_{i=1}^{N}} = 15 \times N 
\quad \Rightarrow \quad \mathcal{O}(N).
\end{align}
For instance, in a 12-cell deployment with $\numCoreMax = 56$ and $\numVfAccMax = 16$, the full brute-force space contains $\approx{5}\times 10^{12}$ combinations, whereas {\name}’s scheduler evaluates only $180$, yielding orders-of-magnitude reduction.

This reduced search space enables {\name} to adapt allocations in real time using 
Algorithm~\ref{alg:profiling_scheduler}. 
For example, consider the 2$\times$2 {100}\thinspace{MHz} cell in Fig.~\ref{fig:Feecback-LUT}.  
In the single-cell case, {\name} selects $\sysConfigCellIdx{i}{1}{1}{1}{1}$ with a RAF confidence of $p=0.542$ as the most energy-efficient configuration meeting the 3-slot deadline.  
As the number of active cells grows ($N \geq 4$), contention degrades performance and the confidence score drops below the threshold ($p=0.469$), prompting the scheduler to escalate to $\sysConfigCellIdx{i}{3}{3}{1}{1}$, the next entry in $\cellProfileRF{\cellConfigVar_i}$, thereby restoring deadline compliance.  
A similar behavior is observed for 4$\times$4 {100}\thinspace{MHz} cells in Fig.~\ref{figs:MOO--4-4-100MHz}. While $\sysConfigCellIdx{i}{4}{4}{4}{4}$ is sufficient for a single cell, multi-cell contention necessitates escalation to $\sysConfigCellIdx{i}{5}{5}{5}{5}$ to maintain real-time guarantees.  
%
Its effectiveness is further demonstrated in the evaluation (Sec.\ref{sec:evaluation}, Fig.\ref{fig:Heter-Cell-Heter-CR}).

\begin{algorithm}[!t]
\caption{\name's multi-cell Scheduler}
\label{alg:profiling_scheduler}
\small
\begin{algorithmic}[1]
\Require $N$ cells with features $\{\cellConfigVar_i\}_{i=1}^{N}$;
         RAF classifier $f(\cdot)$ returning feasibility confidence $[0,1]$;
         learned contention coefficients $(\beta_0,\beta_1)$;
         decision threshold $\tau$.
\Ensure Total compute resource $(\numCoreMax,\numVfAccMax)$ is sufficient to support all $N$ cells to meet the 3-slot processing deadlines; $\bigl|\cellProfileRF{\cellConfigVar_i}\bigr| \ge 1$ for every active cell $i$.

\For{$i=1,2,\dots,N$}
  \State $\cellFeasibleRF{\cellConfigVar_i} \gets \Big\{\, \sysConfigVar_{i}^{(k)} \in \setSysConfigReduced{\numCoreTotal_i}{\numVfAcc_i} ~\Big|~ f(\cellConfigVar_i,\sysConfigVar_{i}^{(k)})\geq \tau \,\Big\}$
  \State $\cellProfileRF{\cellConfigVar_i} \gets \text{sort } \cellFeasibleRF{\cellConfigVar_i} \text{ by } \power(\cdot)$ \Comment{increasing power}
  
  \For{$\sysConfigVar_{i}^{(k)} \in \cellProfileRF{\cellConfigVar_i}$}
     \State obtain feasibility confidence $p_{i}^{k} \gets f(\cellConfigVar_i,\sysConfigVar_{i}^{(k)})$ 
     \State $\hat{p}_{i}^{k} \gets p_{i}^{k} -  (\beta_0 + \beta_1\cdot(N-1))$ 
     \If{$\hat{p} \ge \tau$}
        \State $\sysConfigCellIdx{i}{\numCoreTotalCellIdx{i}}{\numCoreDspCellIdx{i}}{\numCoreAccCellIdx{i}}{\numVfAccCellIdx{i}} \gets \sysConfigVar_{i}^{(k)}$
        \State \textbf{break}
     \EndIf
  \EndFor
\EndFor

\State \textbf{Return} $\left\{ \sysConfigCellIdx{i}{\numCoreTotalCellIdx{i}}{\numCoreDspCellIdx{i}}{\numCoreAccCellIdx{i}}{\numVfAccCellIdx{i}} \right\}_{i=1}^{N}$
\end{algorithmic}
\end{algorithm}

\section{Implementation}
\label{sec:implementation}

\myparatight{Multi-cell implementation.}
We implement {\name} in about 5,000 lines of C++ code, built on top of the open-source {\savannah} codebase~\cite{qi2024savannah}.
Key components include a heterogeneous compute resource scheduler and a set of statically compiled processing pipelines corresponding to different resource allocation strategies. 
At runtime, {\name} reads the active cell configurations and selects the appropriate pre-compiled resource allocation strategy without requiring recompilation or binary modification.

\myparatight{{\namebf} scheduler.}
We implement {\name} scheduler in about 300 lines of Python code, including feature extraction, model fitting, and evaluation.
The model is trained offline using profiling data (detailed in \S\ref{ssec:model-perf}), and at runtime {\name} invokes the scheduler during system initialization to minimize overhead.
Specifically, the main C++ program calls the Python scheduler as a subprocess, interpreting each cell's feature vector $\cellConfigVar$ via JSON input.
The scheduler evaluates $\cellConfigVar$ with a set of $\sysConfigVar$ and outputs the required resource allocation $\sysConfigOpt_i$ for each active cell—i.e., the number of CPU cores and ACC100 VFs needed to satisfy the 3-slot deadline.  
These allocations are returned in JSON format and immediately consumed by the C++ runtime, which binds the specified cores and VFs to each cell’s processing pipeline. Because the scheduler runs only at initialization, all subsequent per-slot scheduling decisions are handled natively in C++, ensuring ultra-low Python overhead during real-time operation.

{\name} also includes over 9K example configuration files that cover a wide range of cell configurations and resource allocation strategies, which are used to generate the profiling datasets for model training and testing.
To support system-specific adaptation, an automated batch script is also provided to perform single-cell profiling across different cell configurations, enabling users to generate training data that matches their hardware environment.

\myparatight{ACC100 virtual function (VF) initialization.}
We integrate two Silicom Lisbon P2 ACC100 accelerators~\cite{acc100} with {\name}, binding both devices to DPDK's \texttt{\small igb\_uio} driver using the \texttt{\small dpdk-devbind} tool.
One ACC100 operates with 16 VFs ($\numVfAccMax=16$) while the second is configured in PF mode to enable direct performance comparisons between VF-based and PF-based configurations.
To manage VF addressability, we enable the I/O Memory Management Unit (IOMMU) in the system BIOS and kernel.
VF configuration parameters, including Queue Groups, DDR memory allocation, and atomic queue settings, are configured through Intel’s {\bbdev} \texttt{\small pf\_bb\_config} tool through customized configuration files.
To support $\sysConfigCellIdx{i}{c}{c-1}{1}{v}$ and $\sysConfigCellIdx{i}{c}{c}{c}{c}$, {\name} utilize DPDK’s threading model and the parallelism of ACC100.
We adopt \texttt{\small rte\_thread} to create RTE threads with proper CPU affinity, resource isolation, and memory-zone consistency under a single EAL environment.
This implementation enables each VF to be attached to a worker core without launching independent EAL instances, supporting up to 16 concurrent VFs per ACC100 hardware accelerator.

\myparatight{Server configuration.}
We perform all profiling (e.g., Fig.~\ref{fig:MOO}) and experiments using two Dell PowerEdge R750 servers, interconnected via a Dell EMC PowerSwitch Z29264F-ON using {100}\thinspace{G} QSFP28 DAC cables.
\emph{Server 1 (\myCodeShort{srv1}) emulates the RUs}, each generating baseband traffic corresponding to multiple UEs. 
It features a 64-core Intel Xeon Gold 6548N CPU @{2.6}\thinspace{GHz} with Ubuntu {24.04.2} LTS and an NVIDIA Mellanox ConnectX-6 Dx dual-port {100}\thinspace{GbE} SmartNIC. 
The traffic streams are aggregated and streamed Server 2.
\emph{Server 2 (\myCodeShort{srv2}) emulates the DUs}, executing the vRAN baseband processing pipeline with {\name}.
It is equipped with a 56-core Intel Xeon Gold 6348 CPU @{2.6}\thinspace{GHz} running Ubuntu {20.04.6} LTS and an NVIDIA Mellanox ConnectX-5 dual-port {100}\thinspace{GbE} NIC. 
To ensure real-time PHY processing, specific CPU cores are isolated and dedicated to {\name}, which are also responsible for interfacing with two Silicom Lisbon P2 ACC100 accelerators via DPDK {21.11.2}.
To minimize context switching and optimize performance, we enable performance mode for the CPU and execute {\name} as a real-time process at the highest scheduling priority on both servers.
We also disable hyper-threading and employ thread affinity, mitigating the impact of sharing resources among logical cores.
\myCodeShort{srv2} also runs {\name}'s scheduling framework and evaluates 
the DL-based and analytical baseline schedulers. 

\section{Evaluation}
\label{sec:evaluation}

\subsection{Experimental Setup}
We deploy {\name} with the emulated RUs and DUs instances on separate servers to evaluate the performance.
\textit{Server-1} emulates the RUs while \textit{Server-2} emulates the DUs.
%
In addition to the worker cores allocated for DSP and ACC100 tasks ($\numCoreDsp$, $\numCoreAcc$), extra CPU cores are reserved for TX/RX streaming and master control on both the RU and DU sides. 
Due to the logging overhead, especially when processing large numbers of cells, the system generates substantial log data; for example, recording 100K frames across sixteen cells produces over {2.0}\thinspace{GB} of log files. This makes it infeasible to sustain detailed logging beyond sixteen cells without impacting system performance.
\emph{However, when logging is disabled, {\name} can support more cells without experiencing processing deferrals, demonstrating its scalability.}
We evaluate various cell configuration $\cellConfigVar(\cellConfigMIMO, \cellConfigBW, \cellConfigTL, \cellConfigTBW, \cellConfigMCS)$.
%
A processing latency deadline of 3-slot ({0.375}\thinspace{ms})
is enforced at the 99.9th percentile across all evaluations.
We report results along two dimensions: (\emph{i}) {\name} single-cell scheduler's performance, and (\emph{ii}) system-level multi-cell performance.

\begin{table}[!t]
\centering
\small
\caption{Random forest performance as a function of forest size. Inference time reported as $\mu$s/$(\cellConfigVar,\sysConfigVar)$ at batch size 1.}
\vspace{-3mm}
\begin{tabular}{rccc}
\toprule
\textbf{\# of Trees} & \textbf{Accuracy} & \textbf{Inference Time} & \textbf{Model Size} \\
\midrule
10   & 98.99\% & 1.42\,$\mu$s & {0.5}\thinspace{MB} \\
50   & 99.10\% & 4.45\,$\mu$s & {2.3}\thinspace{MB} \\
100  & 99.09\% & 4.24\,$\mu$s & {4.4}\thinspace{MB} \\
500  & 99.08\% & 21.30\,$\mu$s & {22.2}\thinspace{MB} \\
1000 & 99.09\% & 43.67\,$\mu$s & {44.4}\thinspace{MB} \\
\bottomrule
\end{tabular}
\vspace{-3mm}
\label{tab:rf-scaling}
\end{table}

\subsection{Single-Cell Scheduler Performance}
\label{ssec:model-perf}

We first evaluate the feasibility predictors used in {\name}, comparing the proposed {\name} single-cell scheduler with six baselines, including DL and analytical models. 
Our evaluation focuses on three metrics:
(\emph{i}) accuracy, which refers to the fraction of test samples for which the model correctly predicts feasibility, i.e., whether the 99.9th percentile latency under a given $(\cellConfigVar,\sysConfigVar)$ defined in \S\ref{ssec:single-cell-opt} meets the 3-slot deadline. (\emph{ii}) inference time per $(\cellConfigVar,\sysConfigVar)$, and (\emph{iii}) model size.

\myparatight{Datasets.}
We collect more than 9K profiling experiments across diverse cell configurations 
$\cellConfigVar(\cellConfigMIMO, \cellConfigBW, \cellConfigTL, \cellConfigTBW, \cellConfigMCS)$ and resource allocation stragety $\sysConfigVar$, 
spanning
\begin{itemize}[leftmargin=*, topsep=2pt, itemsep=1pt]
\item
MIMO dimension, $\cellConfigMIMO \in \{ 1 \times 1 (\text{SISO}), 2 \times 2, 4 \times 4\}$,
\item
Channel BW, $\cellConfigBW \in \{ 100\thinspace\text{MHz}, 200\thinspace\text{MHz}, 400\thinspace\text{MHz} \}$,
\item
Traffic load, $\cellConfigTL \in \{ 25\%, 50\%, 75\%, 100\% \}$,
\item
Transmission BW percentage, $\cellConfigTBW \in \{ 25\%, 50\%, 75\%, 100\% \}$, 
\item
MCS, $\cellConfigMCS \in \{ 10, 12, 17, 20 \}$.
\end{itemize}
For each $(\cellConfigVar,\sysConfigVar)$ pair, we run {\name} for 20K frames and log detailed timing information of all DSP stages. 
This \emph{raw logging} produces more than {300}\thinspace{GB} of trace data.  
From these traces, we extract the 99.9th percentile end-to-end processing latency.  
Each $(\cellConfigVar,\sysConfigVar)$ is labeled as \emph{feasible} (=1) if the latency satisfies the 3-slot deadline, and \emph{infeasible} (=0) otherwise, yielding the initial labeled dataset of 9K samples $(\cellConfigVar,\sysConfigVar)$.

\begin{table}[!t]
\centering
\small
\caption{Deep learning model performance on CPU and GPU. Inference time reported as $\mu$s/$(\cellConfigVar,\sysConfigVar)$ at batch size 1. 
}
\vspace{-3mm}
\begin{tabular}{lcccc}
\toprule
\textbf{Model} & \textbf{Accuracy} & \textbf{\specialcell{ Inference Time \\ (CPU/GPU)}} & \textbf{Model Size} \\
\midrule
MLP  & 99.14\% & 582/743\,$\mu$s & 55.1\,KB \\
1D CNN & 99.26\% & 615/826\,$\mu$s & 52.9\,KB \\
TinyNet & 99.34\% & 570/803\,$\mu$s & 30.1\,KB \\
\bottomrule
\end{tabular}
\vspace{-3mm}
\label{tab:dl-comparison}
\end{table}

To broaden coverage, we further augment the dataset using monotonicity rules observed during profiling:  
(i) if a solution is feasible for $\cellConfigTL$=$a$, then it is also feasible for all $\cellConfigTL$$<a$;  
(ii) if infeasible for $\cellConfigTL$=$a$, then it remains infeasible for all $\cellConfigTL$$\geq a$;  
(iii) the same monotonicity rule applies to $\cellConfigTBW$;  
(iv) if feasible with $c$ CPU cores, then it is also feasible for any $c'>c$.  
Similar rules also stand for $\cellConfigTBW$. Applying these rules expands the dataset to 120{,}596 labeled samples $(\cellConfigVar,\sysConfigVar)$.  
We split the dataset into train/validation/test sets in a 60/20/20 ratio and use the same splits for RAF, DL models, and mathematical baselines to ensure fair comparison.  

\myparatight{Selecting the forest size.}
Table~\ref{tab:rf-scaling} reports the performance of the proposed {\name} single-cell scheduler under different forest sizes.
It can be seen that prediction accuracy only marginally improves with larger forests--from {98.99\%} with 10 trees to {99.10\%} with 50 trees--after which it quickly saturates.
Both the inference time and model size grow linearly with the number of trees.
%
In practice, forests with {50--100} trees provide a favorable tradeoff between accuracy ($\approx${99.10\%}) and efficiency ($\leq${5}\thinspace{$\upmu$s} per inference), and therefore, we empirically select \name single-cell scheduler with 50 trees for the remainder of our evaluations.

\begin{figure}[!t]
    \centering
    \vspace{-3mm}
    \subfloat[SISO, $\sysConfigCellIdx{i}{1}{1}{1}{1}$]{
    \label{figs:SISO-confidence}
    \includegraphics[width=0.50\columnwidth]{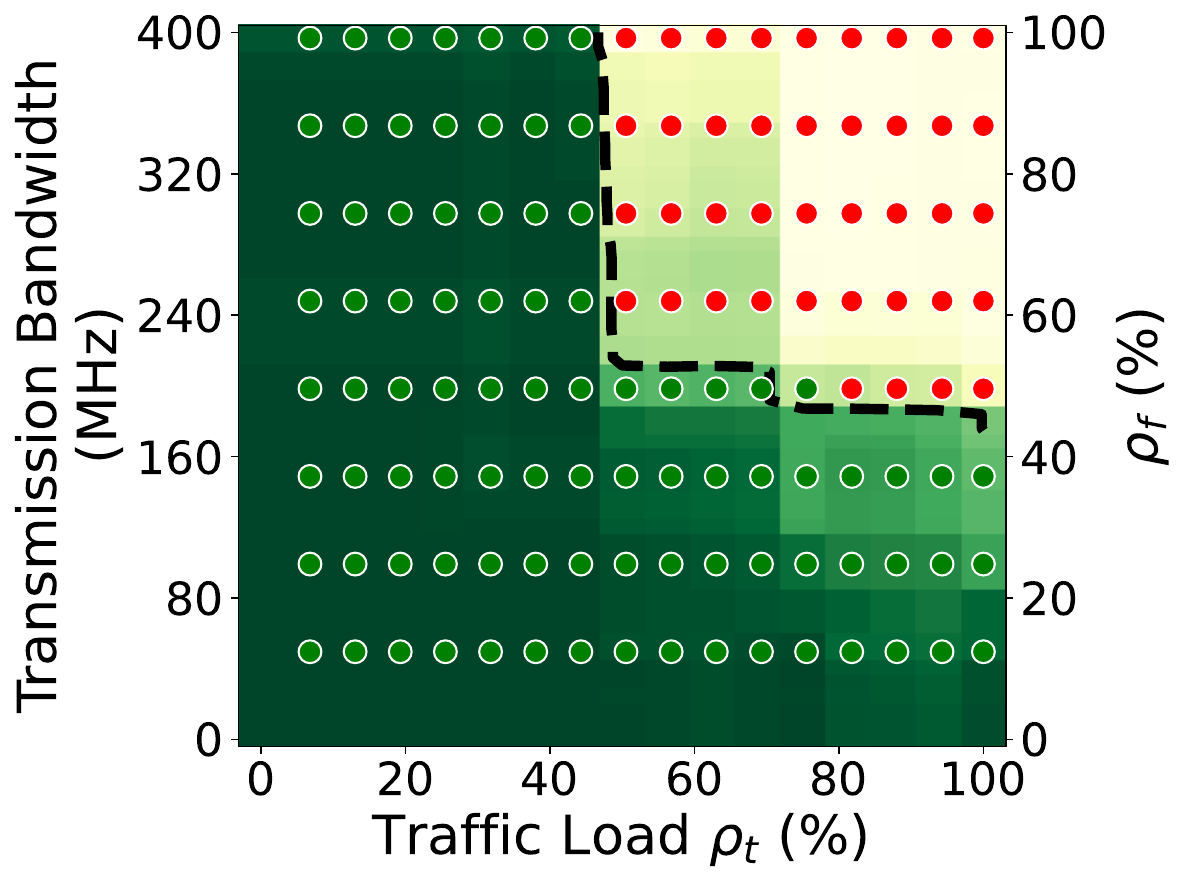}}
    \subfloat[2$\times$2 MIMO, $\sysConfigCellIdx{i}{4}{4}{0}{0}$]{
    \label{figs:2-2-confidence}
    \includegraphics[width=0.50\columnwidth]{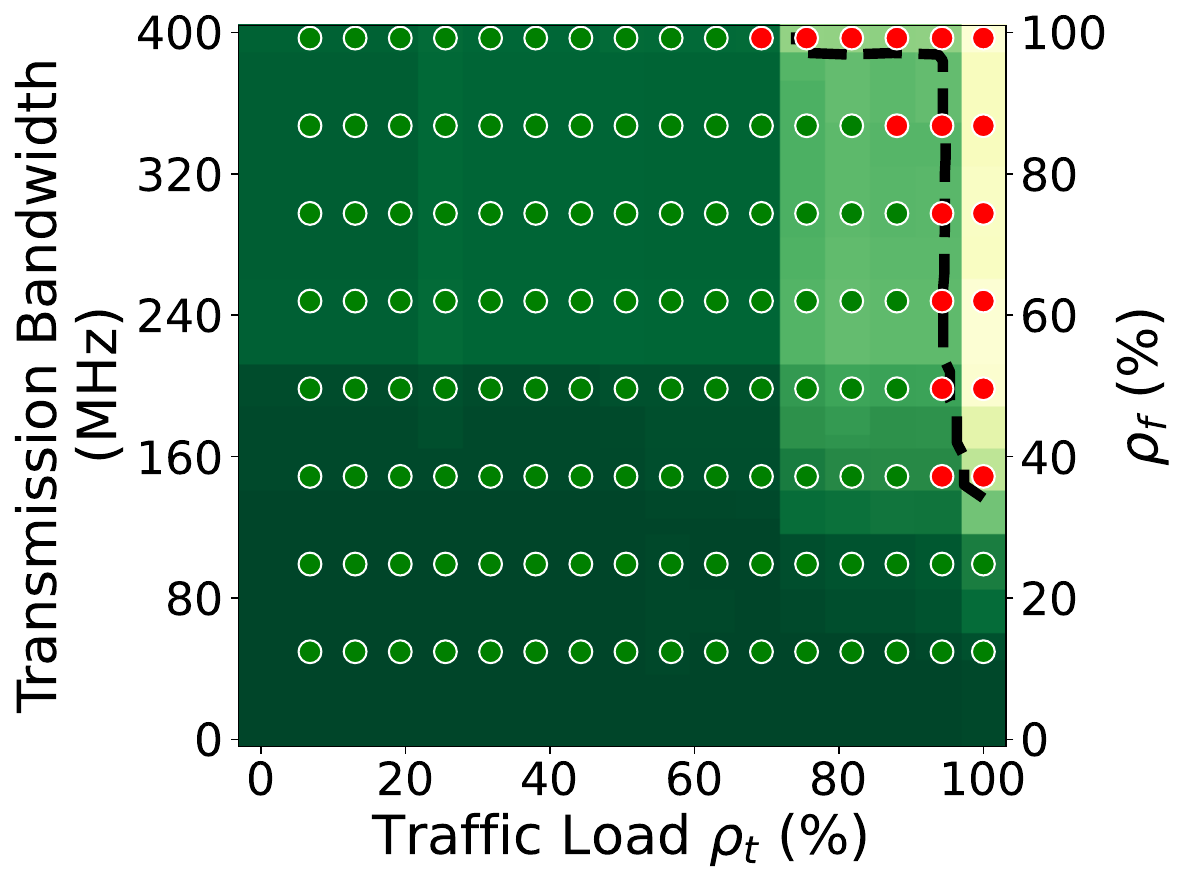}}
    \vspace{-3mm}
    \caption{{\namebf} single-cell scheduler's predictions 
    with $MCS=18$. 
    Confidence heatmaps are thresholded at $\tau=0.5$ (black dashed line).
    Each green/red dot corresponds to measured 99.9\textsuperscript{th} processing latency outcomes from executing {\namebf} over 20K frames.}
    \label{fig:rf-unseen}
    \vspace{-5mm}
\end{figure}

\begin{figure*}[!t]
    \centering
    \includegraphics[width=1.8\columnwidth]{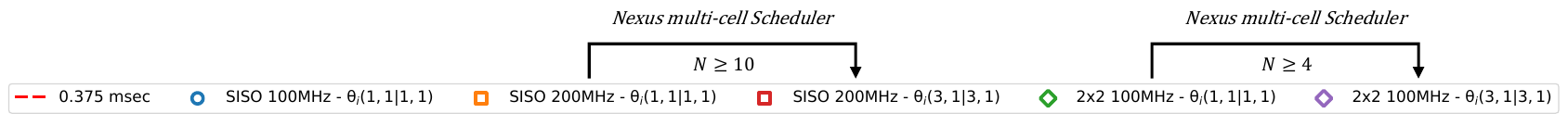}
    
    \vspace{-5mm}
    \subfloat[Resource allocation based on single-cell Profiling]{
    \includegraphics[width=0.98\columnwidth]{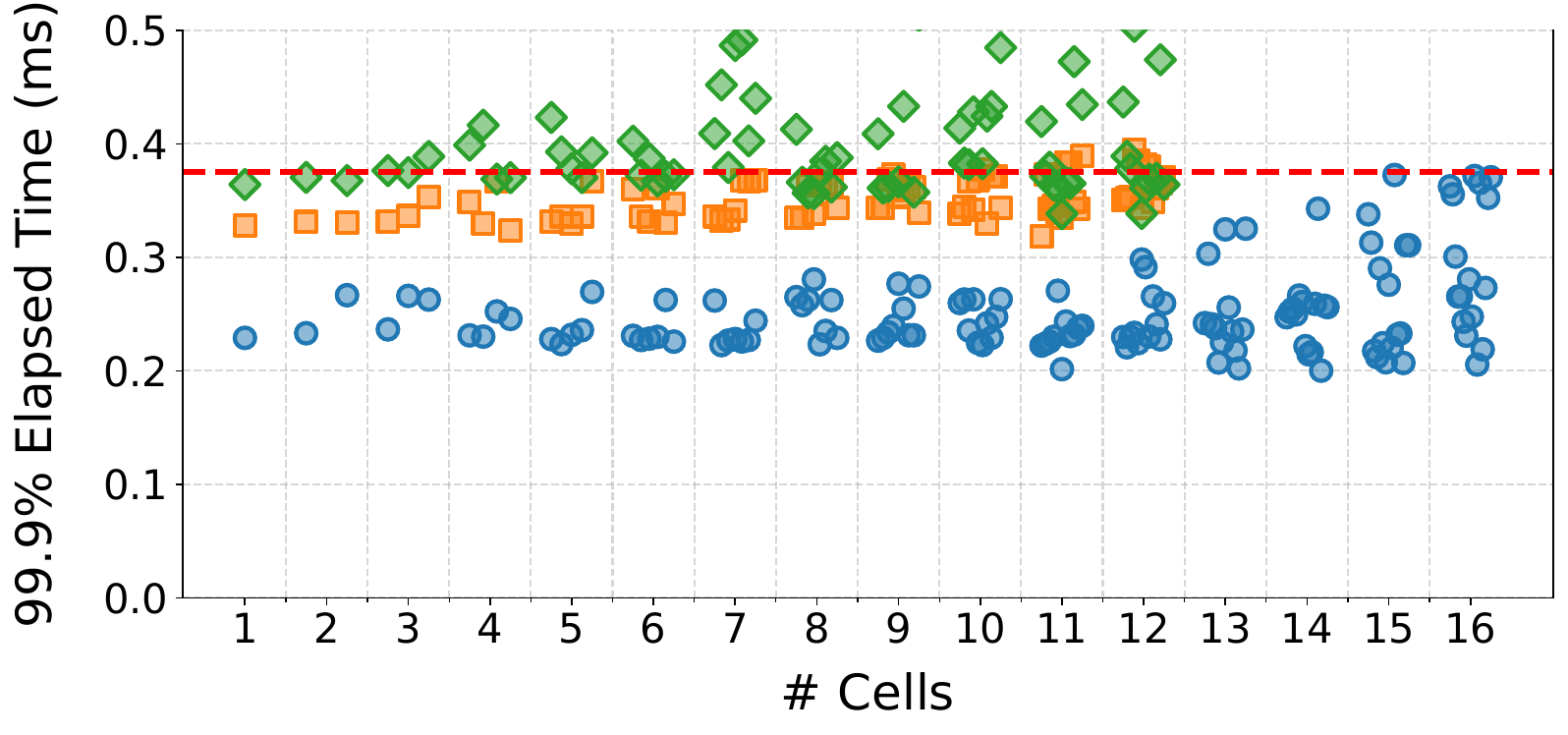}
    \label{fig: Same-Cell-MCS10}}
    \hspace{5mm}
    \subfloat[Resource allocation based on {\name} multi-cell scheduler]{
    \includegraphics[width=0.98\columnwidth]{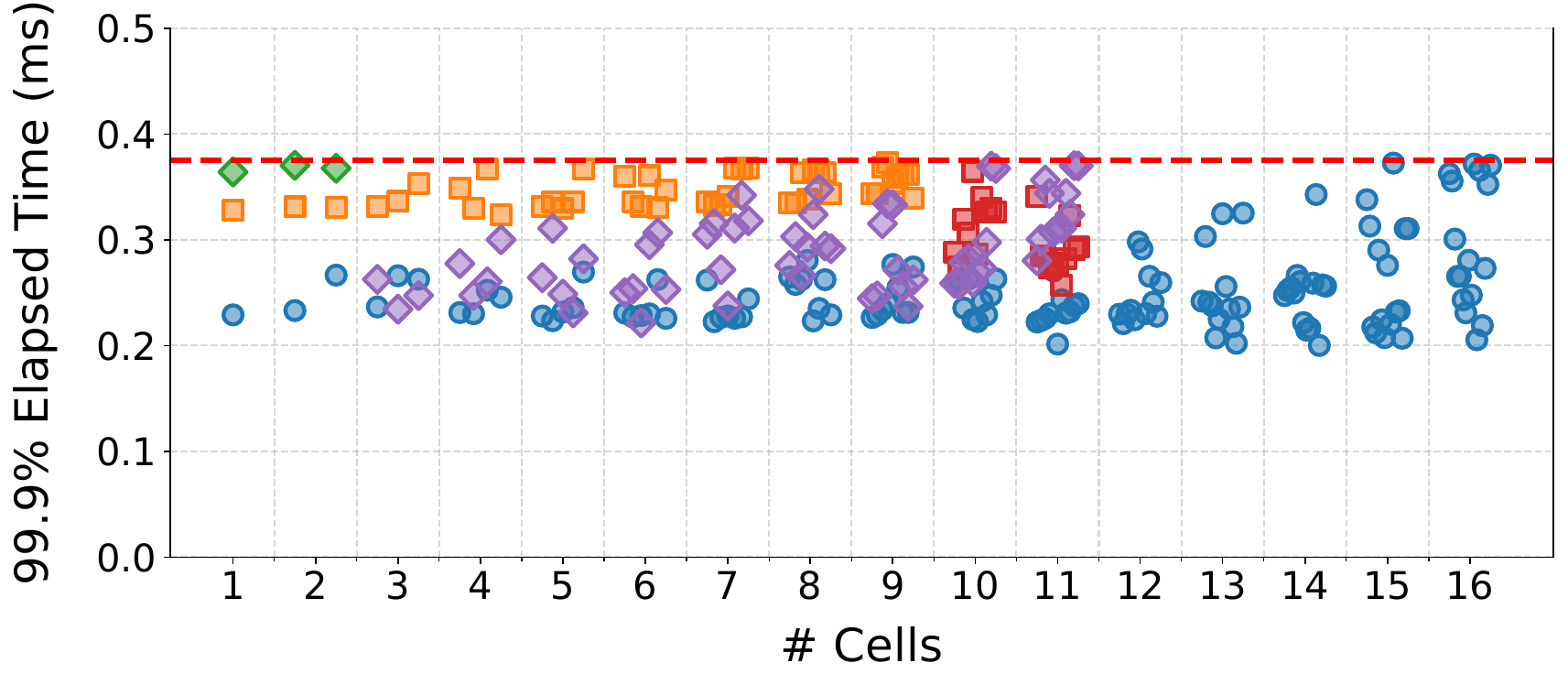}
    \label{fig: Same-Cell-MCS17}}
    \vspace{-3mm}
    \caption{99.9\textsuperscript{th} percentile processing latency vs. the number of concurrently running cells for different cell configurations. 
    Each group of points shows latency measurements for $N$ cells using the same configuration. 
    In (b), for 2$\times$2 {100}\thinspace{MHz} cells with $\sysConfigCellIdx{i}{1}{1}{1}{1}$, latency exceeds the deadline when $N \geq 4$, prompting {\namebf} to escalate to $\sysConfigCellIdx{i}{3}{3}{1}{1}$, restoring compliance. 
    }
    \label{fig:Feecback-LUT}
    \vspace{-3mm}
\end{figure*}
\myparatight{Comparison with DL models.}
Table~\ref{tab:dl-comparison} compares {\name} single-cell scheduler against three DL models: an MLP~\cite{popescu2009multilayer, singh2019study} with three fully connected layers (128–64–32 neurons with ReLU activations), a 1D CNN~\cite{azizjon20201d} with two convolutional layers (32 and 64 filters, kernel size 3) followed by a dense layer (128 neurons), and a TinyNet~\cite{TinyNet, xu2022etinynet} variant with two depthwise-separable convolutional layers (16 and 32 filters) followed by global average pooling and a dense layer. 
The three DL models are trained with the same training/validation dataset and evaluated on the same test dataset as {\name} single-cell scheduler to ensure a fair comparison, and they are evaluated on both CPU and GPU.
All DL models achieve comparable accuracy ($\approx${99.20\%}), but their inference latency is two orders of magnitude higher than RAF.
While GPU execution improves throughput, it introduces non-negligible CPU–GPU transfer overhead, 
which is why GPU inference in our experiments is often slower than CPU inference.
One trade-off is that {\name} single-cell scheduler consumes more memory than compact DL models, since each tree stores explicit branching nodes and split thresholds rather than compressed weight matrices.
For example, a 50-tree RAF occupies about $2.2$\,MB, compared to $\leq 1$\,MB for the DL models.
However, this overhead is negligible in modern servers equipped with hundreds of gigabytes of memory and is easily justified by {\name} single-cell scheduler's significantly lower inference latency and CPU-only deployment.

We further examine sensitivity to input feature dimensionality.
For {\name} single-cell scheduler with 50 trees, increasing the input from 4 to 6 features only increases inference latency from ${3.09}\,\mu$s to ${3.16}\,\mu$s ($+2.4\%$). By contrast, DL inference time grows substantially: for a 1D CNN model, increasing from 4 to 6 features increases latency from ${613}\,\mu$s to ${636}\,\mu$s ($+3.8\%$).
While the relative percentage increases appear comparable, the absolute overhead is dramatically different: {\name} single-cell scheduler incurs an additional ${0.07}\,\mu$s per $(\cellConfigVar,\sysConfigVar)$, whereas the DL model incurs ${23}\,\mu$s.
This highlights that RAF inference is effectively insensitive to feature dimensionality, whereas DL inference grows noticeably as features increase due to deeper/wider layers and heavier matrix multiplications.
These results directly validate the rationale for adopting RAF outlined in \S~\ref{ssec:RAF-model}. 

\myparatight{Comparison with mathematical baselines.}
We also compare the {\name} single-cell scheduler against three lightweight analytical baselines: 
(\emph{i}) \emph{logistic regression}, a linear classifier trained on the same features as RAF and DL; 
(\emph{ii}) \emph{threshold-based heuristics}, which determine feasibility by applying cutoffs on $\cellConfigTL$, $\cellConfigBW$, and $\cellConfigTBW$ learned directly from profiling. 
Specifically, for each feature we identify the maximum feasible value observed in the training set and use these thresholds as decision rules;  
and
(\emph{iii}) \emph{fixed-zone rules}, which define static feasible regions by combining such thresholds across features and assigning infeasible otherwise.  
All models are evaluated on the same test set, which ensures that comparisons are consistent and fair.
%
While these analytical baselines are computationally trivial, they perform poorly due to their inability to capture non-linear feature interactions.  
Logistic regression achieves 93.2\% accuracy, and the threshold-based model reaches 92.7\%.  
The fixed-zone rule performs much worse, collapsing to near-random accuracy (42\%).  
These results highlight that such simple models fail to capture the coupling between PHY-layer features, whereas {\name}'s RAF scheduler achieves near-DL accuracy with substantially higher reliability.
RAF thus strikes a favorable balance, achieving an accuracy ($\approx 99.10\%$) close to that of the DL models at much faster inference time.

\myparatight{Generalization to unseen cell configurations.}
We next evaluate whether {\name} single-cell scheduler can generalize beyond our dataset by testing on unseen cell configurations. 
Fig.~\ref{fig:rf-unseen} shows two examples: 
Fig.~\ref{figs:SISO-confidence} represents a SISO cell with MCS18 using $\sysConfigCellIdx{i}{1}{1}{1}{1}$, and Fig.~\ref{figs:2-2-confidence} represents a 2$\times$2 MIMO cell with MCS18 using $\sysConfigCellIdx{i}{4}{4}{0}{0}$, both evaluated under varying $\cellConfigTL$ and $\cellConfigTBW$. 
In these maps, green regions denote configurations predicted to be feasible (meeting the 3-slot deadline), while white regions denote infeasible allocations. The green and red dots correspond to measured ground-truth outcomes from 20K frames (feasible/infeasible, respectively).
It can be seen that {\name} single-cell scheduler achieves strong prediction accuracy of overall $98.8\%$ even for these unseen configurations, with close agreement between predicted feasible regions and measured outcomes. By contrast, traditional models such as threshold-based heuristics fail to capture these non-linear feasibility boundaries, yielding accuracies below {90\%}.
These results confirm that {\name} single-cell scheduler is not only accurate and efficient, but also robust in extrapolating to new cell scenarios.

\subsection{Multi-Cell Scheduler Performance}
All system-level evaluations are conducted at 100\% $\cellConfigTL$, 100\% $\cellConfigTBW$ and $\cellConfigMCS=17$ to stress-test {\name} under extreme conditions as MCS17 imposes a higher computational load~\cite{ding2020agora, qi2024savannah}.

\myparatight{Multi-cell with identical cell configurations and resource allocations.}
$\sysConfig{1}{1}{1}{1}$ can reliably meet the 3-slot processing deadline on a single core based on {\name} scheduler for a cell with configurations 1$\times$1 {100}\thinspace{MHz}, 1$\times$1 {200}\thinspace{MHz}, or 2$\times$2 {100}\thinspace{MHz}.
%
We begin by evaluating how many such cells can be simultaneously supported by {\name} under extreme conditions.
Fig.~\ref{fig:Feecback-LUT} shows the results when {\name}  simultaneously serves $N = \{1, 2, \dots, 16\}$ cells with identical configurations, each handled by $\sysConfigCellIdx{i}{1}{1}{1}{1}$. 

Fig.~\ref{fig:Feecback-LUT} shows that {\name} scales when concurrently serving multiple cells.
%
As illustrated in Fig.~\ref{fig: Same-Cell-MCS10}, when using $\sysConfigVar_{i}^{(k)}$ directly from the single-cell prediction result, for 1$\times$1 {100}\thinspace{MHz} cells, {\name} supports up to sixteen simultaneous cells using a total number of sixteen worker cores and sixteen ACC100 VFs, resulting in aggregated data rate of {3.91}\thinspace{Gbps}. However, 1$\times$1 {200}\thinspace{MHz} and 2$\times$2 {200}\thinspace{MHz} cells present corner cases. As we are running ten 1$\times$1 {200}\thinspace{MHz} cells or four 2$\times$2 {100}\thinspace{MHz} cells, their single-cell confidence rate is already near the threshold under $\sysConfig{1}{1}{1}{1}$, and additional parallel cells introduce contention that causes deadline violations (see \S{\ref{ssec:scheduler}}).
In response, {\name} would transition to the next $\sysConfigVar_{i}^{(k)}$ in $\mathcal{P}_{\mathrm{RAF}}$, which is {$\sysConfigCellIdx{i}{3}{3}{1}{1}\ \forall i$ for those cells.
As shown in Fig.~\ref{fig: Same-Cell-MCS17},
by adjusting resource allocation to $\sysConfigCellIdx{i}{3}{3}{1}{1}$ for each cell, {\name} supports up to eleven cells for both 1$\times$1 {200}\thinspace{MHz} and 2$\times$2 {100}\thinspace{MHz} while meeting the 3-slot processing deadline, with a total number of 33 worker cores and 11 ACC100 VFs, resulting in an aggregated data rate of {5.37}\thinspace{Gbps}.
Our experiments are limited to eleven cells, which already occupy 55 of the 56 available cores (including streaming and master cores).
%
\begin{figure}[!t]
    \centering
    \hspace{0.5cm} \includegraphics[width=0.98\columnwidth]{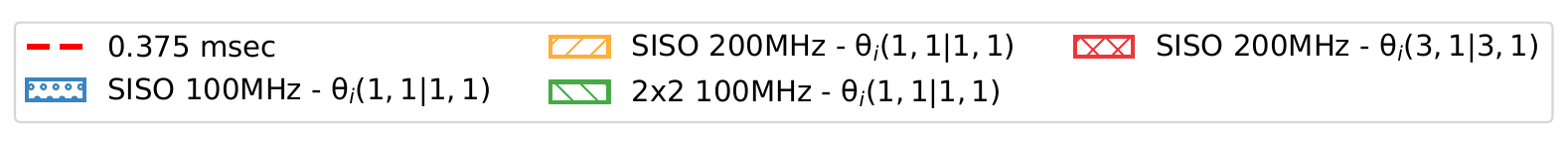}
    
    \vspace{-4mm}
    \subfloat{
    \includegraphics[width=0.90\columnwidth]{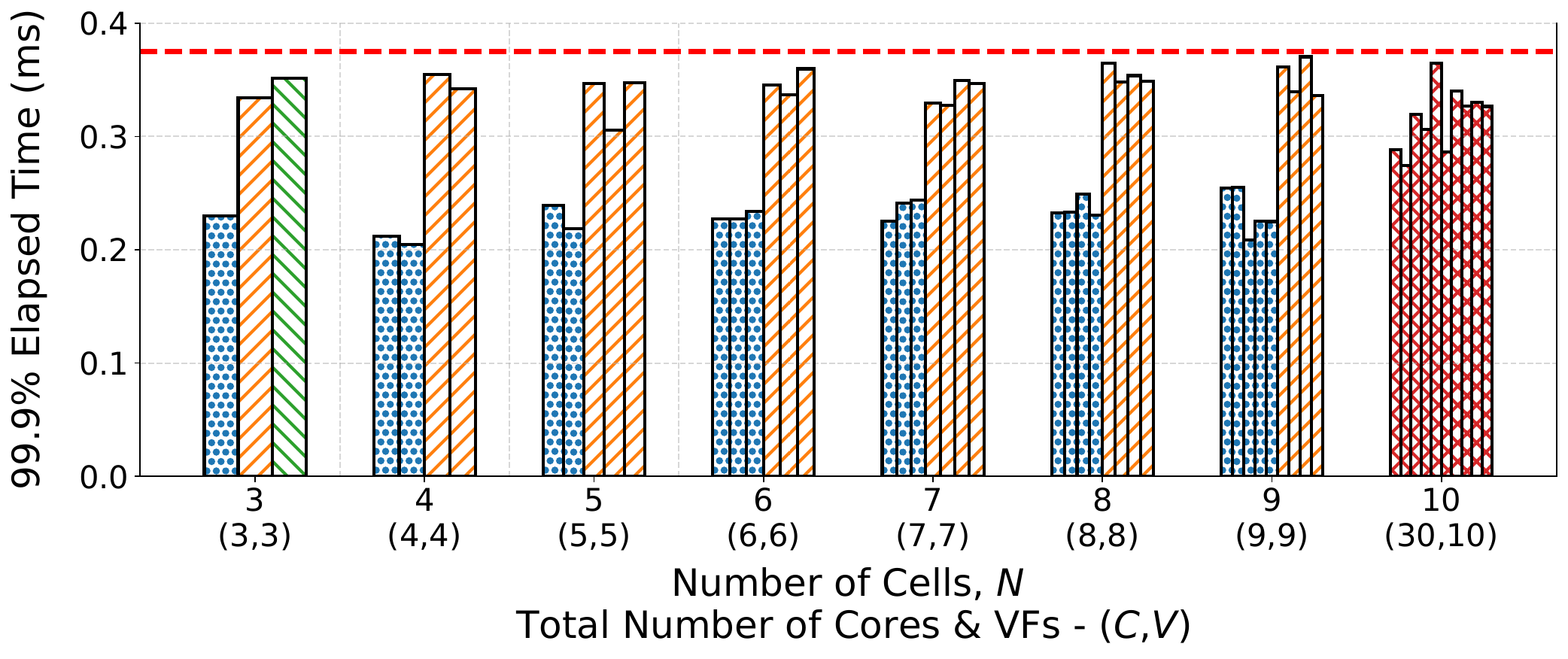}}
    \vspace{-3mm}
    \caption{99.9\textsuperscript{th} percentile processing latency for heterogeneous cell configurations processed concurrently using the same resource allocation. 
    }
    \label{fig:Heter-Cell-Same-CR}
    \vspace{-3mm}
\end{figure}
%
It is worth noting that even when the same allocation strategy appears unchanged across different $N$, the multi-cell scheduler is still actively evaluating feasibility through RAF confidence. 
For example, in the case of 1$\times$1 {100}\thinspace{MHz} cells with $\sysConfigCellIdx{i}{1}{1}{1}{1}$, the RAF confidence decreases from $p{=}0.87$ at $N{=}1$ to $p{=}0.53$ at $N{=}16$. 
Although contention reduces the margin, the confidence remains above the threshold $\tau$, so the allocation is considered feasible for all sixteen cells. 

\myparatight{Multi-cell with heterogeneous cell configurations and identical resource allocations.}
Next, we evaluate {\name}'s ability to handle heterogeneous cells while employing identical resource allocations for each cell.
Fig.~\ref{fig:Heter-Cell-Same-CR} shows the 99.9th-percentile processing latency when {\name} concurrently serves $N=\{3, 4, 5, 6, 7, 8, 9, 10\}$ cells with $\sysConfigCellIdx{i}{1}{1}{1}{1}$ or $\sysConfigCellIdx{i}{3}{3}{1}{1}$.
The results for $N$ cells have $N$ bars, corresponding to the 99.9th-percentile processing latency for each of the $N$ cells.
%
When running nine cells--five 1$\times$1 {100}\thinspace{MHz} cells and four 1$\times$1 {200}\thinspace{MHz} cells--{\name} processes all frames within the 3-slot deadline, achieving a total data rate of {3.17}\thinspace{Gbps} using overall nine worker cores and nine ACC100 VFs, simultaneously supporting heterogeneous cell configurations.

\begin{figure}[!t]
    \centering
    \hspace{0.3cm} 
    \includegraphics[width=0.90\columnwidth]{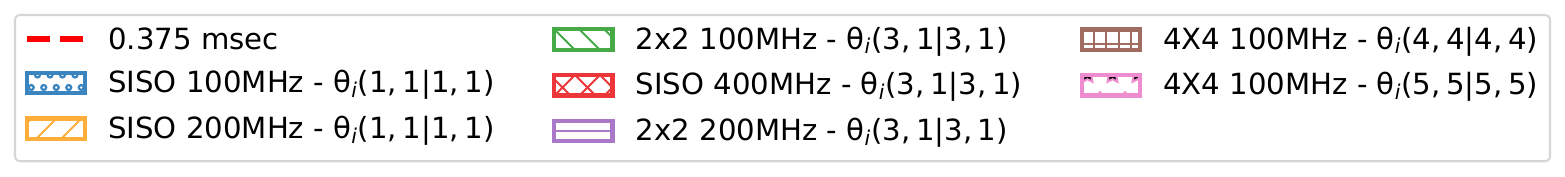}
    
    \vspace{-4mm}
    \subfloat{
    \includegraphics[width=0.95\columnwidth]{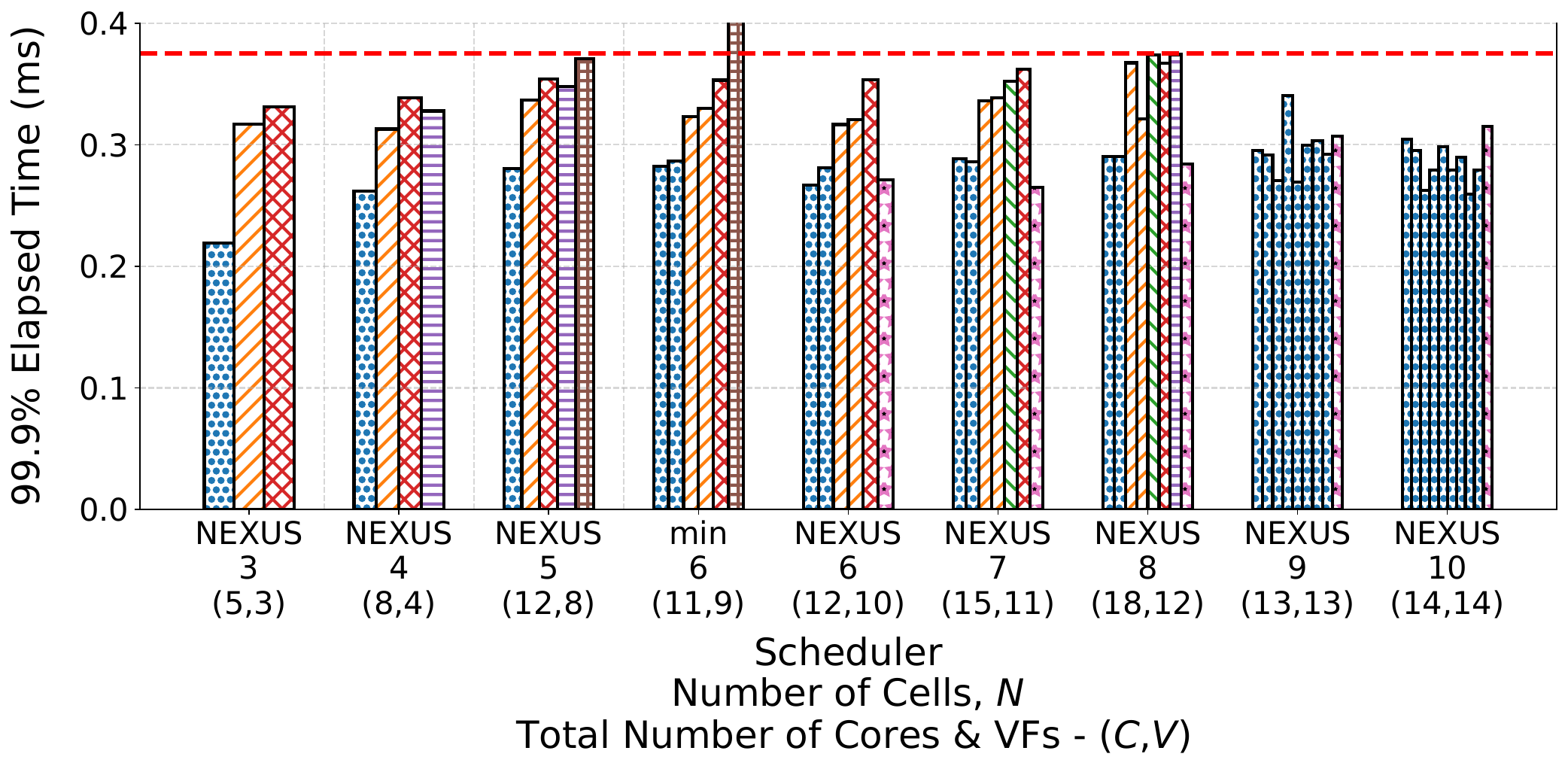}}
    \hspace{2mm}
    \vspace{-3mm}
    \caption{99.9\textsuperscript{th} percentile processing latency for heterogeneous cells with heterogeneous resource allocation. 
    {\namebf} represents the scheduler proposed in this work, while \textit{min} represents using resources based on single-cell profiling. 
    }
    \label{fig:Heter-Cell-Heter-CR}
    \vspace{-5mm}
\end{figure}

\myparatight{Multi-cell with heterogeneous cell configurations and heterogeneous resource allocations.}
Finally, we evaluate the performance of {\name} in scenarios with heterogeneous cell configurations that use different processing techniques.
Fig.~\ref{fig:Heter-Cell-Heter-CR} shows the 99.9th-percentile processing time for individual cells when {\name} concurrently serves $N = \{3, 4, 5, 6, 7, 8, 9, 10\}$ heterogeneous cells.
%
When configured to serve five cells with a 5-th 4$\times$4 {100}\thinspace{MHz} cell with high complexity, {\name} selects $\sysConfigCellIdx{5}{4}{4}{4}{4}$ to handle this cell based on Algorithm~\ref{alg:profiling_scheduler}.
Fig.~\ref{fig:Heter-Cell-Heter-CR} shows all five cells with a total data rate of {2.68}\thinspace{Gbps} using 12 worker cores and eight ACC100 VFs are able to meet the 3-slot deadline. 

For the scenario with six cells, 
the prediction confidence for a 4$\times$4 {100}\thinspace{MHz} cell processing latency under resource allocation $\sysConfigCellIdx{6}{4}{4}{4}{4}$ was estimated to be infeasible based on {\eqref{eq:rf_confidence_adjusted}}
and is experimentally confirmed in Fig.~\ref{fig:Heter-Cell-Heter-CR} using the \textit{min} scheduler across six cells. 
In response, {\name} invokes its multi-cell scheduler and adjusts the resource allocation for this cell to $\sysConfigCellIdx{6}{5}{5}{5}{5}$.  
With this updated resource allocation, the latency is improved to {0.268}\thinspace{ms}.
As a result, the corresponding bar for the 4$\times$4 {100}\thinspace{MHz} cell appears lower in the six-cell case than in the five-cell case in Fig.~\ref{fig:Heter-Cell-Heter-CR}. 
This adaptive resource reassignment enables all six cells to meet their real-time processing deadlines, achieving a total data rate of {3.41}\thinspace{Gbps} using 12 worker cores and 10 ACC100 VFs. 
%
%
The results highlight {\name}'s ability to scale to highly heterogeneous deployments while maintaining real-time guarantees.


\section{Conclusion}

In this paper, we presented {\name}, the first system to enable efficient virtualized, multi-cell mmWave PHY baseband processing on a single heterogeneous compute server. {\name} addresses the growing need for efficient computational resource sharing in vRAN systems. We introduced detailed mechanisms for sharing the Intel ACC100 accelerator across multiple cores and Virtual Functions, supported by a RAF-based scheduler that minimizes power consumption while meeting real-time deadlines. Our evaluation demonstrates that {\name} supports up to sixteen concurrent cells with a combined data rate of {5.37}\thinspace{Gbps} and more when logging is disabled across diverse deployment scenarios. By bridging the gap between baseband computation and resource virtualization, {\name} opens the door to scalable and energy-efficient vRAN architectures.

\begin{acks}
This work was supported in part by NSF grants CNS-2211944, AST-2232458, CNS-2330333, and CNS-2443137.
This work was also supported in part by the Center for Ubiquitous Connectivity (CUbiC), sponsored by Semiconductor Research Corporation (SRC) and Defense Advanced Research Projects Agency (DARPA) under the JUMP 2.0 program.
We thank Anuj Kalia (Microsoft), Rahman Doost-Mohammady and Andrew Sedlmayr (Rice University), and Danyang Zhuo (Duke University) for their contributions to this work.
\end{acks}

\appendix
\bibliographystyle{ACM-Reference-Format}
\bibliography{reference}

\end{document}